\documentclass[aps,prl,amsmath,amssymb,twocolumn,superscriptaddress,showpacs]{revtex4-1}
\usepackage{color}
\usepackage{amsmath,amssymb}
\usepackage{graphicx}	
\mathchardef\mhyphen="2D

\def\be{\begin{eqnarray}}   
\def\ee{\end{eqnarray}}
\def\vecb{\boldsymbol}

\begin{document}

\author{Shunsuke~A.~Sato}
\email{ssato@ccs.tsukuba.ac.jp}
\affiliation 
{Center for Computational Sciences, University of Tsukuba, Tsukuba 305-8577, Japan}
\affiliation 
{Max Planck Institute for the Structure and Dynamics of Matter, Luruper Chaussee 149, 22761 Hamburg, Germany}

\author{Hideki~Hirori}
\affiliation
{Institute for Chemical Research, Kyoto University, Uji, Kyoto 611-0011, Japan}

\author{Yasuyuki~Sanari}
\affiliation
{Institute for Chemical Research, Kyoto University, Uji, Kyoto 611-0011, Japan}

\author{Yoshihiko~Kanemitsu}
\affiliation
{Institute for Chemical Research, Kyoto University, Uji, Kyoto 611-0011, Japan}

\author{Angel~Rubio}
\email{angel.rubio@mpsd.mpg.de}
\affiliation 
{Max Planck Institute for the Structure and Dynamics of Matter, Luruper Chaussee 149, 22761 Hamburg, Germany}
\affiliation 
{Center for Computational Quantum Physics (CCQ), Flatiron Institute, 162 Fifth Avenue, New York, NY
10010, USA}

\title{High-order harmonic generation in graphene: nonlinear coupling of intra and interband transitions}

\begin{abstract}
We investigate high-order harmonic generation (HHG) in graphene with a quantum master equation approach. The simulations reproduce the observed enhancement in HHG in graphene under elliptically polarized light [N.~Yoshikawa \textit{et al}, Science 356, 736 (2017)]. On the basis of a microscopic decomposition of the emitted high-order harmonics, we find that the enhancement in HHG originates from an intricate nonlinear coupling between the intraband and interband transitions that are respectively induced by perpendicular electric field components of the elliptically polarized light. Furthermore, we reveal that contributions from different excitation channels destructively interfere with each other. This finding suggests a path to potentially enhance the HHG by blocking a part of the channels and canceling the destructive interference through band-gap or chemical potential manipulation.
\end{abstract}

\maketitle


High-order harmonic generation (HHG) is an extreme photon-upconversion process based on highly nonlinear light-matter interactions. HHG was originally observed in atomic gas systems more than thirty years ago \cite{McPherson:87,Ferray_1988}. Several years later, the microscopic mechanism underlying HHG in noble gases was beautifully explained by a simple semi-classical model, the so-called three-step model \cite{kulander1993dynamics,PhysRevLett.71.1994,PhysRevA.49.2117}. On the practical side, high coherence in these upconversion processes allows us to generate extremely short light pulses, presenting a novel avenue to time-domain investigations of ultrafast electron dynamics in matter \cite{RevModPhys.72.545,RevModPhys.81.163,Goulielmakis2010,Beck_2014,Schultze1348,Lucchini916,Zurch2017,Siegrist2019,Volkov2019}.

Since the first observation of HHG in solids by Gimire \textit{et al.} \cite{Ghimire2011}, HHG in solids has been attracting much interest as it may have various applications ranging from the development of novel light sources \cite{Ghimire2019} to probing of microscopic information of matter \cite{PhysRevLett.115.193603,chacon2018observing,Silva2019}. So far, experimental studies on HHG in solids have explored various materials \cite{Schubert2014,Luu2015,You2017,Yoshikawa736,PhysRevB.102.041125}, and the theoretical aspects of HHG in solids have been intensively investigated with various approaches \cite{PhysRevB.77.075330,PhysRevLett.113.073901,PhysRevB.91.064302,PhysRevA.97.011401,PhysRevA.95.043416,PhysRevLett.118.087403}.

Recently, attosecond transient absorption spectroscopy and microscopic simulations clarified that nonlinear coupling of intraband and interband transitions play significant roles in ultrafast modification of optical properties and in nonlinear photocarrier-injection processes \cite{Lucchini916,Schlaepfer2018,buades2018attosecond}. Hence the nonlinear coupling of the two kinds of transitions may be a key to accessing the microscopic physics behind a light-induced phenomenon and may offer a novel opportunity to control it. While intraband and interband transitions have been discussed in the context of HHG in solids \cite{PhysRevB.77.075330,golde2009microscopic}, still detailed roles of nonlinear coupling of these transitions have not yet been investigated.

Yoshikawa \textit{et al.} recently reported that the HHG in graphene can be enhanced by elliptically-polarized light \cite{Yoshikawa736}. This observation is distinct from the HHG in noble gases, where HHG is significantly suppressed with an increase in the ellipticity of light \cite{PhysRevA.48.R3437,PhysRevA.50.R3585,Liang_1994}. Therefore, HHG in graphene under elliptically-polarized light would offer an opportunity to look into the microscopic mechanism underlying HHG in solids. However, the microscopic mechanism of the HHG enhancement under elliptically-polarized light has been still unclear. 

In this work, we investigate the enhancement of HHG in graphene with elliptically-polarized light, by employing a quantum master equation with a simple two-band model. The simple modeling of electron dynamics in graphene fairly captures the experimentally-observed enhancement of HHG and provides a microscopic insight into the mechanism. The model indicates a significant role of the nonlinear coupling between light-induced intraband and interband transitions in the enhancement of HHG, demonstrating a destructive interference among multiple HHG channels.

To describe the electron dynamics, we employ a quantum master equation with a two-band approximation for the Dirac cone of graphene \cite{PhysRevB.99.214302,sato_njp_2019}. In the model, the time propagation of the one-body reduced density matrix at each Bloch wavevector $\vecb k$ is described by
\be
\frac{d}{dt}\rho_{\vecb k}(t) = \frac{1}{i\hbar}\left [
H_{\vecb k + \vecb A(t)}, \rho_{\vecb k} (t) 
\right ] + \hat D\left [ \rho_{\vecb k}(t) \right ],
\label{eq:quantum-master}
\ee
where $H_{\vecb k + \vecb A(t)}$ is the Hamiltonian, and $\hat D[\rho_{\vecb k}(t)]$ is a relaxation operator. In this work, we employ the following $2$-by-$2$ Hamiltonian matrix:
\be
H_{\vecb k + \vecb A(t)} &=& v_F\tau_z \sigma_x \left [
k_x + A_x(t)
\right] +
v_F \sigma_y \left [
k_y + A_y(t)
\right] 
+ \frac{\Delta}{2}\sigma_z, \nonumber \\
\label{eq:dirac-ham}
\ee
where $\sigma_{j}$ are Pauli matrices, $k_{j}$ are the $j$-component of the Bloch wavevector $\vecb k$, and $A_{j}(t)$ is the $j$-component of the vector potential $\vecb A(t)$, which corresponds to the applied electric fields, $\vecb E(t)=-\dot {\vecb A}(t)$. The band-gap $\Delta$ is set to zero for graphene unless stated otherwise. Here, $\tau_z$ determines the chirality of the system (either $+1$ or $-1$). We evaluate observables as the average of two calculations with opposite chiralities. We set the Fermi velocity $v_F$ to $1.12\times 10^6$~m/s in accordance with an \textit{ab-initio} simulation \cite{PhysRevLett.101.226405}. The relaxation operator $\hat D [\rho_{\vecb k}(t)]$ is constructed by making the relaxation-time approximation with a longitudinal relaxation time of $T_1=100$~fs and transverse relaxation time of $T_2=20$~fs \cite{PhysRevB.99.214302,sato_njp_2019,SuppMat}. Note that the relaxation operator also depends on the chemical potential $\mu$; $\mu$ is set to zero (charge neutrality point) unless stated otherwise.

To describe the applied electric fields, we employ the following form of the vector potentials
\be
\vecb A(t) = &-&\frac{cE_{0,x}}{\omega_0} \vecb e_x \cos \left (\omega_0 t - \frac{\pi}{4}\right )
\cos^4\left (\frac{\pi}{T_{full}}t \right ) \nonumber \\
&-&\frac{cE_{0,y}}{\omega_0} \vecb e_y \cos \left (\omega_0 t + \frac{\pi}{4}\right )
\cos^4\left (\frac{\pi}{T_{full}}t \right )
\label{eq:laser-pulse}
\ee
in the domain $-T_{full}/2<t<T_{full}/2$; the potential is zero outside this domain. In accordance with the experimental conditions in Ref.~\cite{Yoshikawa736}, we set the mean photon energy $\hbar \omega_0$ to $260$~meV. The full pulse duration $T_{full}$ is set to $100$~fs. We will investigate the electron dynamics by changing the peak field strength of the applied laser fields, $E_{0,x}$ and $E_{0,y}$.

Employing a time-dependent density matrix, $\rho_{\vecb k}(t)$, we can evaluate the induced electric current as
\be
\vecb J(t, E_{0,x}, E_{0,y}) = \frac{1}{(2\pi)^2}\int d\vecb k
\mathrm{Tr}\left[ \hat {\vecb J}_{\vecb k} \rho_{\vecb k}(t)\right ],
\label{eq:bare-current}
\ee
where $\hat {\vecb J}$ is the current operator defined by
\be
\vecb{\hat{J}}_{\vecb k}(t) = -\frac{\partial H_{\vecb k+\vecb A(t)}}{\partial \vecb{A}(t)}.
\ee
Note that the current defined in Eq.~(\ref{eq:bare-current}) depends on $E_{0,x}$ and $E_{0,y}$ via $\vecb A(t)$ in Eq.~(\ref{eq:laser-pulse}), and for clarity we will indicate this dependence in the next equation by using the notation $\vecb J(t, E_{0,x}, E_{0,y})$.

In experiments, HHG occurs not only at the center of the beam-spot but also on the whole focal area. To make our model more realistic, we employ the following intensity-averaging procedure to approximate the results for the case of a Gaussian beam profile \cite{PhysRevA.97.011401,SuppMat}:
\be
\vecb J^{ave}(t) = \int^1_0 d \alpha \frac{1}{\alpha}\vecb J(t,\alpha E_{0,x}, \alpha E_{0,y}).
\label{eq:focal-average}
\ee

The power spectrum of the high-order harmonics polarized along the $j$-direction can be evaluated with the current as
\be
I_{j}(\omega) \sim \omega^2 \left |\int dt J^{ave}_{j}(t)e^{i\omega t} \right |^2,
\ee
where $J^{ave}_{j}(t)$ is the $j$-component of the current vector $\vecb J^{ave}(t)$. Furthermore, the intensity of the $n$th-order harmonics can be evaluated by integrating the power spectrum within a finite range as
\be
I^{n{\rm th}}_{j} = \int^{n\omega_0+\frac{1}{2}\omega_0}_{n\omega_0-\frac{1}{2}\omega_0}
d\omega  I_{j}(\omega).
\ee

\begin{figure}[htbp]
  \includegraphics[width=0.9\columnwidth]{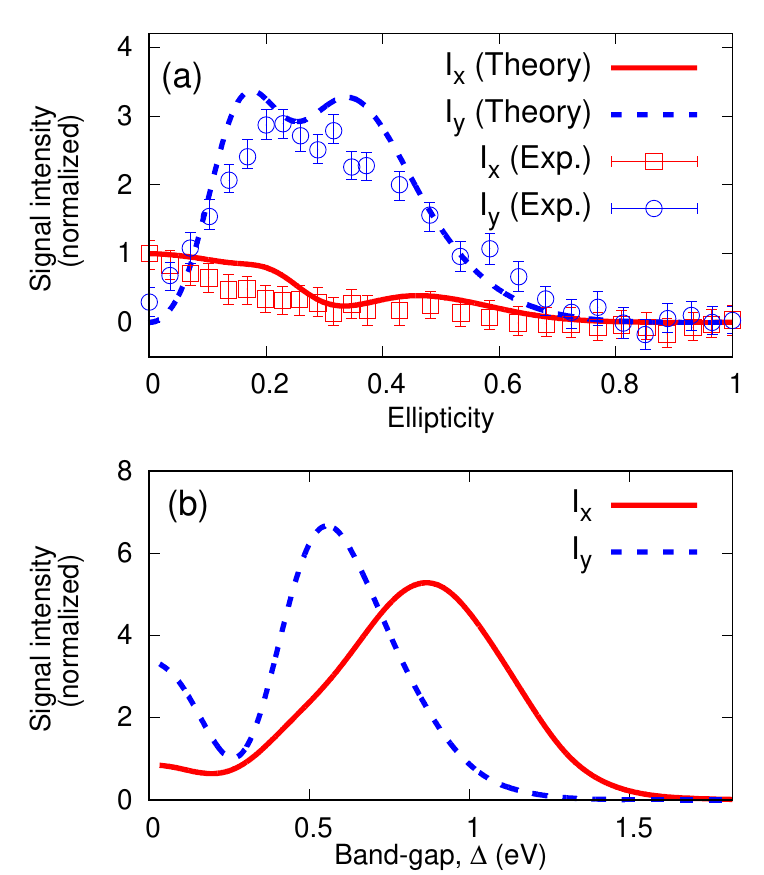}
\caption{\label{fig:graphene_ellipt_gap_hhg_7th}
The 7th-order harmonic intensity from graphene under elliptically polarized light: (a) harmonic intensity as a function of ellipticity, $E_{0,y}/E_{0,x}$. The experimental data \cite{Yoshikawa736} are also shown: (b) harmonic intensity as a function of the band-gap $\Delta$. The harmonic intensity is decomposed into the two polarization axes: the major $I^{7th}_x$ (red-solid) and minor $I^{7th}_y$ (blue-dotted) axes of the driving elliptically polarized light. 
}
\end{figure}

First, we evaluate the ellipticity dependence of the HHG by fixing the peak field strength $\sqrt{E^2_{0,x}+E^2_{0,y}}$ to $6.5$~MV/cm inside the material. The major axis of the elliptically polarized light is set to the $x$-axis while the minor axis is set to the $y$-axis. Figure~\ref{fig:graphene_ellipt_gap_hhg_7th}~(a) shows the signal intensity of the 7th-order harmonics as a function of laser ellipticity, $E_{0,y}/E_{0,x}$. The intensity $I^{7th}_{j}$ is separately computed for the different polarization directions ($j=x$ or $j=y$) of the emitted harmonics. As seen from the figure, when the applied laser field is linearly polarized in the $x$-direction, the emitted high harmonics are also linearly polarized in the $x$-direction. Once the applied fields become elliptical, the emitted harmonics also become elliptical, having both $x$ and $y$-components. Interestingly, the $y$-component $I^{7th}_y$ rapidly increases with the increase in driver ellipticity and becomes much larger than the $x$-component $I^{7th}_x$, demonstrating enhancement in HHG in graphene by elliptically-polarized light. Once the driver ellipticity further increases and approaches one (circularly-polarized light), the emitted harmonics is significantly suppressed due to the circular symmetry of the Dirac cone. These observations are consistent with the experimental results \cite{Yoshikawa736}. Thus, it has been demonstrated that the simple Dirac band model with the relaxation-time approximation contains sufficient ingredients to describe the HHG in graphene and can be used to investigate its microscopic origin.

To obtain further insight into the phenomena, we evaluated the harmonic intensity by changing the band-gap $\Delta$. Figure~\ref{fig:graphene_ellipt_gap_hhg_7th}~(b) shows the 7th-order harmonic intensity as a function of band-gap $\Delta$. Here, we used the same field strength as in Fig.~\ref{fig:graphene_ellipt_gap_hhg_7th}~(a). The ellipticity is set to $0.17$, by the harmonic intensity is maximized at $\Delta=0$. Surprisingly, the harmonic intensity can be significantly enhanced by increasing the band-gap. Furthermore, the $x$-component of the harmonic intensity $I^{7th}_x$ shows a peak around a band-gap of $0.8$~eV, which is close to the energy of three photons ($0.78$~eV), while the $y$-component $I^{7th}_y$ shows a peak around a band-gap of $0.5$~eV, which is close to the energy of two photons ($0.52$~eV). The enhancement and formation of peaks that occur as the band-gap increases indicate that multi-photon processes play a significant role in HHG in graphene, while Zener tunneling is expected to have only a minor contribution in the present regime.

The enhancement in HHG with the increase in band-gap $\Delta$ may be regarded as a counter-intuitive consequence because an increase in the gap tends to block a part of the transitions. In fact, the high-order harmonics vanish once the band-gap becomes significantly large, as shown in Fig.~\ref{fig:graphene_ellipt_gap_hhg_7th}~(b). To understand the enhancement in HHG with the increase in the gap, we propose a microscopic mechanism based on destructive interference between multiple channels: high-order harmonics are generated as a superposition of multiple signals from various microscopic paths due to nonlinear coupling of intraband and interband transitions. We further suppose that the multiple signals may destructively interfere with each other, and the total signal may be weakened. When such destructive interference plays a significant role, HHG may be enhanced by increasing the gap, because in so doing contributions can be partly suppressed, and the destructive interference can be canceled.

\begin{figure}[htbp]
  \includegraphics[width=0.98\columnwidth]{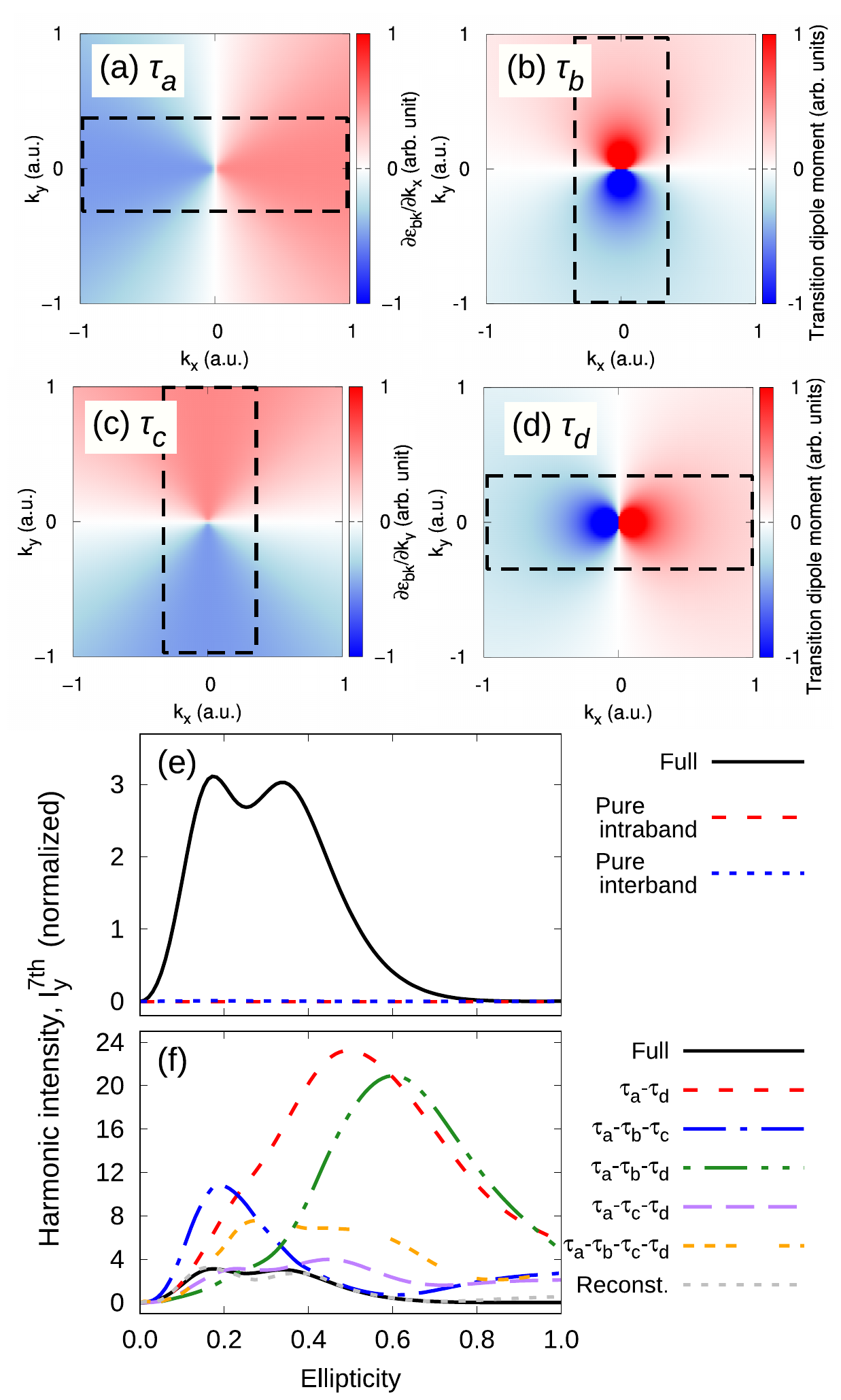}
\caption{\label{fig:graphene_ellipt_intra_inter_hhg_7th}
(a-d) Strength profiles of intraband and interband transitions in $k$-space. The origin is set to the Dirac point. Strong interaction areas are indicated by black dashed squares. (e) The 7th-order harmonic intensity with the full intraband and interband transitions (black solid line), solely with the intraband transitions (red dashed line), and solely with the interband transitions (blue dotted line). (f) The 7th-order harmonic intensity with various decomposed transitions.
}
\end{figure}

To examine our hypothesis, let us investigate the interference of different HHG contributions from the viewpoint of intraband and interband transitions. Here, we partly turn off the transitions based on the instantaneous eigenbasis representation, where the intraband transitions appear in the diagonal elements of the Hamiltonian and the interband transitions appear in the off-diagonal elements \cite{PhysRevB.98.035202,SuppMat}. Elliptically-polarized light consists of two polarization components, and each of them induces intraband and interband transitions. Hence, we can consider the four kinds of light-induced transitions. For later convenience, we label them as follows: intraband transitions induced by the $x$-component of the electric fields $(\tau_{a})$ and by the $y$-component $(\tau_{b})$; likewise, interband transitions induced by the $x$-component of the electric fields $(\tau_{c})$ and by the $y$-component $(\tau_{d})$. Figures~\ref{fig:graphene_ellipt_intra_inter_hhg_7th}~(a-d) show the strength distributions of each transition in $k$-space. Here, the strength of the intraband transitions is evaluated as the gradient of the single-particle energy, $\partial \epsilon_{b,\vecb k}/\partial \vecb k$, because the main contribution from the intraband transitions is the modulation of the dynamical phase factor, $\exp \left [-i\int^t dt' \epsilon_{b,\vecb k + \vecb A(t')} \right ]$. The strength of interband transitions is evaluated by the transition dipole moment. As seen from Figs.~\ref{fig:graphene_ellipt_intra_inter_hhg_7th}~(a) and (b), the intraband and interband transitions induced by the $x$-component of the electric fields have alternating strength distributions in $k$-space; when one transition becomes stronger, the other becomes weaker. On the other hand, as seen from Figs.~\ref{fig:graphene_ellipt_intra_inter_hhg_7th}~(a) and (d), the intraband and interband transitions induced respectively by the perpendicular components of the electric fields have similar strength distribution.

To elucidate the roles of the intraband and interband transitions in HHG, we can compute the electron dynamics by turning off part of them. In Fig.~\ref{fig:graphene_ellipt_intra_inter_hhg_7th}~(e), the 7th-order harmonic intensity for the $y$-direction, $I^{7th}_y$, with both intraband and interband transitions is shown as the black-solid line. The results for only intraband transitions are shown as the red dashed line, while those for only interband transitions are shown as the blue dotted line. Here, we have used the same conditions as in Fig.~\ref{fig:graphene_ellipt_gap_hhg_7th}~(b) except the band-gap $\Delta$, which is set to a small value of $0.035$~eV to avoid numerical singularity in the intraband-interband transition analysis at the Dirac point. As the figure clearly shows, neither pure intraband nor interband transitions can induce the HHG. Therefore, HHG originates from a nonlinear coupling of interband and interband transitions. 

To study the roles of the coupling among the intraband and interband transitions, we can decompose the current $\vecb J^\mathrm{ave}(t)$ into the coupling components of the transitions as follows:
\be
\vecb{J}^{\mathrm{ave}}(t)&=&\sum_{\tau} \vecb{J}^\mathrm{ave}_{\tau}(t)
+\sum_{\left \{\tau, \sigma\right \}} \vecb{J}^\mathrm{ave}_{\tau,\sigma}(t) \nonumber \\
&+&\sum_{\left \{\tau, \sigma, \delta \right \}} \vecb{J}^\mathrm{ave}_{\tau,\sigma, \delta}(t)
+\vecb{J}^\mathrm{ave}_{\tau_a, \tau_b, \tau_c, \tau_d}(t),
\label{eq:current-decomp}
\ee
where $\tau$, $\sigma$, and $\delta$ denote the labels of the transitions ($\tau_a$, $\tau_b$, $\tau_c$, and $\tau_d$). In Eq.~(\ref{eq:current-decomp}), there are $15$ terms with four kinds of current: $\vecb{J}^\mathrm{ave}_{\tau}(t)$ is current induced solely by the transition $\tau$. $\vecb{J}^\mathrm{ave}_{\tau,\sigma}(t)$ is current induced by the coupling of two transitions, $\tau$ and $\sigma$. Likewise, $\vecb{J}^\mathrm{ave}_{\tau,\sigma,\delta}(t)$ is the current induced by the coupling among three transitions, $\tau$, $\sigma$, and $\delta$. Finally, $\vecb{J}^\mathrm{ave}_{\tau_a, \tau_b, \tau_c, \tau_d}(t)$ is the current induced by the coupling of all four transitions. For more details, see Supplemental Material~\cite{SuppMat}.

We evaluated the high-order harmonic intensity with the decomposed currents in Eq.~(\ref{eq:current-decomp}) instead of the total current $\vecb J^{ave}(t)$. Figure~\ref{fig:graphene_ellipt_intra_inter_hhg_7th}~(f) shows the $7$th-order harmonic intensity $I^{7th}_y$ as a function of ellipticity for various decomposed current. Here, only the five major contributions, $\vecb{J}^\mathrm{ave}_{\tau_a,\tau_d}(t)$, $\vecb{J}^\mathrm{ave}_{\tau_a,\tau_b,\tau_c}(t)$, $\vecb{J}^\mathrm{ave}_{\tau_a,\tau_b,\tau_d}(t)$, $\vecb{J}^\mathrm{ave}_{\tau_a,\tau_c,\tau_d}(t)$, $\vecb{J}^\mathrm{ave}_{\tau_a,\tau_b,\tau_c}(t)$ and $\vecb{J}^\mathrm{ave}_{\tau_a,\tau_b,\tau_c,\tau_d}(t)$, are shown, while all the other contributions are rather minor~\cite{SuppMat}. In fact, the reconstructed signal from the five major contributions (grey dotted line) shows fair agreement with the full signal (black solid line). Remarkably, all the decomposed results in Fig.~\ref{fig:graphene_ellipt_intra_inter_hhg_7th}~(b) have a larger harmonic intensity than the total signal. Hence the results clearly demonstrate destructive interference among the various contributions. Therefore, our hypothesis, i.e., destructive interference of HHG, is clearly supported by the theoretical results.

As seen from Fig.~\ref{fig:graphene_ellipt_intra_inter_hhg_7th}~(f), the coupling of the intraband transition induced by the $x$-component of the electric fields $(\tau_a)$ and the intraband transitions induced by the $y$-component of the fields $(\tau_d)$ shows the largest contribution to the harmonic intensity (red-dashed line). Therefore, the cross-coupling between the intraband and interband transitions induced respectively by the perpendicular components of the electric fields plays a significant role in the enhancement of HHG in graphene under elliptically-polarized light. In fact, all five major contributions include cross-coupling of the intraband and interband transitions with the perpendicular field components. This observation can be straightforwardly understood in terms of the transition strength distribution in Figs.~\ref{fig:graphene_ellipt_intra_inter_hhg_7th}~(a)-(d). Under linearly-polarized light, the induced intraband and interband transitions have alternating strength distributions in $k$-space; when one transition becomes stronger, the other becomes weaker. Hence the coupling of the intraband and interband transitions is expected to be weak under linearly polarized light. In contrast, under elliptically-polarized light, the intraband and interband transitions induced respectively by the perpendicular components of light have similar strength distributions; when one transition becomes stronger, the other also becomes stronger. Hence, the coupling of the intraband and interband transitions becomes stronger, resulting in an enhancement in HHG in graphene under elliptically polarized light.

\begin{figure}[htbp]
  \includegraphics[width=0.9\columnwidth]{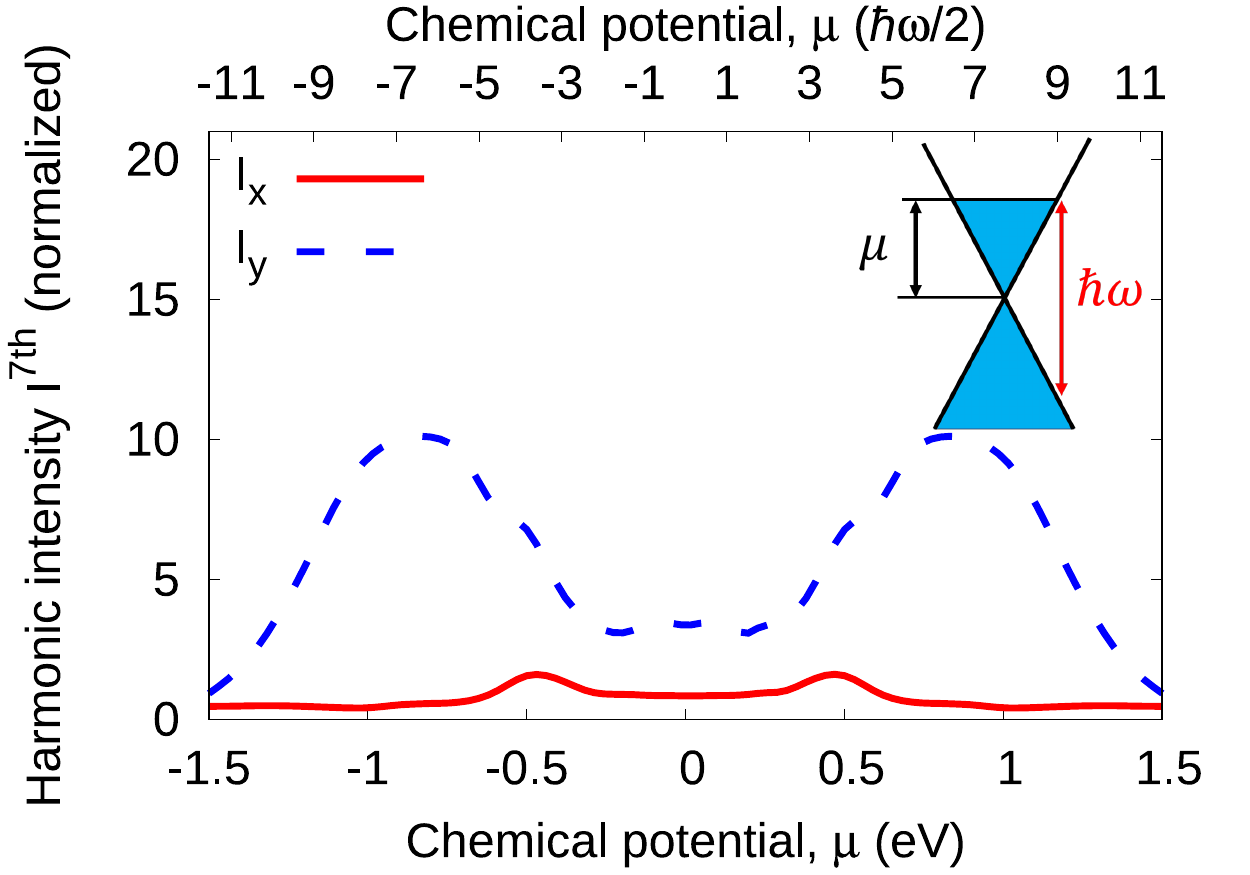}
\caption{\label{fig:graphene_chempot_hhg_7th}
The chemical potential dependence of the 7th-order harmonic intensity for a given ellipticity $E_{0,y}/E_{0,x} = 0.17$.
}
\end{figure}

Having established the destructive interference mechanism of HHG in solids, we propose a way to enhance HHG by canceling the destructive interference with the chemical potential shift. Since the chemical potential shift suppresses a part of the transitions in graphene by Pauli blocking (see the inset of Fig.~\ref{fig:graphene_chempot_hhg_7th}), the destructive interference of multiple channels may be canceled by the tuning of the chemical potential, and this should result in an enhancement of HHG. To demonstrate the impact of the chemical potential shift, we evaluate the 7th-order harmonic intensity from graphene by changing the chemical potential $\mu$. Figure~\ref{fig:graphene_chempot_hhg_7th} shows the 7th-order harmonic intensity as a function of the chemical potential $\mu$. Here, we used the same field strength as Fig.~\ref{fig:graphene_ellipt_gap_hhg_7th}~(b) and set the ellipticity to $0.17$. As shown in Fig.~\ref{fig:graphene_chempot_hhg_7th}, the high-order harmonic intensity can be enhanced by tuning the chemical potential. Hence the proposed method of enhancement in the HHG based on the destructive interference mechanism has clearly been demonstrated. Furthermore, the harmonic intensity shows the peak around $\left |\mu \right |\sim 1~\mathrm{eV}\sim 7\hbar \omega/2$. Note that, once the absolute value of the chemical potential reaches half of the $n$-th-photon energy, $|\mu|\sim n\hbar \omega_0/2$, the $n$-photon resonant processes are suppressed by the Pauli blocking. Thus, the enhancement in HHG and the peak feature in Fig.~\ref{fig:graphene_chempot_hhg_7th} further indicate the significance of multi-photon processes in the present regime.

In conclusion, we investigated and identified the microscopic mechanism underlying HHG in graphene exposed to elliptically polarized light by employing the quantum master equation with the simple Dirac cone \cite{PhysRevB.99.214302,SuppMat}. We found that the nonlinear coupling between the intraband and interband transitions is the microscopic origin of the HHG in graphene. In particular, the cross-coupling between the intraband and interband transitions induced by the perpendicular components of the electric fields causes the enhancement in HHG under elliptically polarized light, reflecting the unique transition strength profiles of graphene in $k$-space, as shown in Figs.~\ref{fig:graphene_ellipt_intra_inter_hhg_7th}~(a-d). Our findings of the interference effects on HHG will lead to a general understanding of HHG in solids, and also provide novel techniques to increase up the HHG efficiency. Indeed, we have demonstrated that the high-order harmonics can be enhanced by tuning the chemical potential $\mu$ or the band-gap $\Delta$, by blocking a part of the transitions. Furthermore, the interference mechanism may open a way to control the high-order harmonic generation with a large degree of freedom by tuning phases of the HHG channel contributions and by flipping the destructive interference to constructive interference through optimization of the applied laser fields.


\begin{acknowledgments}
This work was supported by JSPS KAKENHI Grant Numbers 20K14382 and 19H05465, the European Research Council (ERC-2015-AdG694097), the International Collaborative Research Program of Institute for Chemical Research, Kyoto University (grant \# 2020-13), the Cluster of Excellence 'Advanced Imaging of Matter' (AIM), Grupos Consolidados (IT1249-19) and SFB925 "Light induced dynamics and control of correlated quantum systems". The Flatiron Institute is a division of the Simons Foundation.
\end{acknowledgments}

\bibliography{ref}

\begin{thebibliography}{41}%
\makeatletter
\providecommand \@ifxundefined [1]{%
 \@ifx{#1\undefined}
}%
\providecommand \@ifnum [1]{%
 \ifnum #1\expandafter \@firstoftwo
 \else \expandafter \@secondoftwo
 \fi
}%
\providecommand \@ifx [1]{%
 \ifx #1\expandafter \@firstoftwo
 \else \expandafter \@secondoftwo
 \fi
}%
\providecommand \natexlab [1]{#1}%
\providecommand \enquote  [1]{``#1''}%
\providecommand \bibnamefont  [1]{#1}%
\providecommand \bibfnamefont [1]{#1}%
\providecommand \citenamefont [1]{#1}%
\providecommand \href@noop [0]{\@secondoftwo}%
\providecommand \href [0]{\begingroup \@sanitize@url \@href}%
\providecommand \@href[1]{\@@startlink{#1}\@@href}%
\providecommand \@@href[1]{\endgroup#1\@@endlink}%
\providecommand \@sanitize@url [0]{\catcode `\\12\catcode `\$12\catcode
  `\&12\catcode `\#12\catcode `\^12\catcode `\_12\catcode `\%12\relax}%
\providecommand \@@startlink[1]{}%
\providecommand \@@endlink[0]{}%
\providecommand \url  [0]{\begingroup\@sanitize@url \@url }%
\providecommand \@url [1]{\endgroup\@href {#1}{\urlprefix }}%
\providecommand \urlprefix  [0]{URL }%
\providecommand \Eprint [0]{\href }%
\providecommand \doibase [0]{http://dx.doi.org/}%
\providecommand \selectlanguage [0]{\@gobble}%
\providecommand \bibinfo  [0]{\@secondoftwo}%
\providecommand \bibfield  [0]{\@secondoftwo}%
\providecommand \translation [1]{[#1]}%
\providecommand \BibitemOpen [0]{}%
\providecommand \bibitemStop [0]{}%
\providecommand \bibitemNoStop [0]{.\EOS\space}%
\providecommand \EOS [0]{\spacefactor3000\relax}%
\providecommand \BibitemShut  [1]{\csname bibitem#1\endcsname}%
\let\auto@bib@innerbib\@empty
\bibitem [{\citenamefont {McPherson}\ \emph {et~al.}(1987)\citenamefont
  {McPherson}, \citenamefont {Gibson}, \citenamefont {Jara}, \citenamefont
  {Johann}, \citenamefont {Luk}, \citenamefont {McIntyre}, \citenamefont
  {Boyer},\ and\ \citenamefont {Rhodes}}]{McPherson:87}%
  \BibitemOpen
  \bibfield  {author} {\bibinfo {author} {\bibfnamefont {A.}~\bibnamefont
  {McPherson}}, \bibinfo {author} {\bibfnamefont {G.}~\bibnamefont {Gibson}},
  \bibinfo {author} {\bibfnamefont {H.}~\bibnamefont {Jara}}, \bibinfo {author}
  {\bibfnamefont {U.}~\bibnamefont {Johann}}, \bibinfo {author} {\bibfnamefont
  {T.~S.}\ \bibnamefont {Luk}}, \bibinfo {author} {\bibfnamefont {I.~A.}\
  \bibnamefont {McIntyre}}, \bibinfo {author} {\bibfnamefont {K.}~\bibnamefont
  {Boyer}}, \ and\ \bibinfo {author} {\bibfnamefont {C.~K.}\ \bibnamefont
  {Rhodes}},\ }\href {\doibase 10.1364/JOSAB.4.000595} {\bibfield  {journal}
  {\bibinfo  {journal} {J. Opt. Soc. Am. B}\ }\textbf {\bibinfo {volume} {4}},\
  \bibinfo {pages} {595} (\bibinfo {year} {1987})}\BibitemShut {NoStop}%
\bibitem [{\citenamefont {Ferray}\ \emph {et~al.}(1988)\citenamefont {Ferray},
  \citenamefont {L'Huillier}, \citenamefont {Li}, \citenamefont {Lompre},
  \citenamefont {Mainfray},\ and\ \citenamefont {Manus}}]{Ferray_1988}%
  \BibitemOpen
  \bibfield  {author} {\bibinfo {author} {\bibfnamefont {M.}~\bibnamefont
  {Ferray}}, \bibinfo {author} {\bibfnamefont {A.}~\bibnamefont {L'Huillier}},
  \bibinfo {author} {\bibfnamefont {X.~F.}\ \bibnamefont {Li}}, \bibinfo
  {author} {\bibfnamefont {L.~A.}\ \bibnamefont {Lompre}}, \bibinfo {author}
  {\bibfnamefont {G.}~\bibnamefont {Mainfray}}, \ and\ \bibinfo {author}
  {\bibfnamefont {C.}~\bibnamefont {Manus}},\ }\href {\doibase
  10.1088/0953-4075/21/3/001} {\bibfield  {journal} {\bibinfo  {journal}
  {Journal of Physics B: Atomic, Molecular and Optical Physics}\ }\textbf
  {\bibinfo {volume} {21}},\ \bibinfo {pages} {L31} (\bibinfo {year}
  {1988})}\BibitemShut {NoStop}%
\bibitem [{\citenamefont {Kulander}\ \emph {et~al.}(1993)\citenamefont
  {Kulander}, \citenamefont {Schafer},\ and\ \citenamefont
  {Krause}}]{kulander1993dynamics}%
  \BibitemOpen
  \bibfield  {author} {\bibinfo {author} {\bibfnamefont {K.}~\bibnamefont
  {Kulander}}, \bibinfo {author} {\bibfnamefont {K.}~\bibnamefont {Schafer}}, \
  and\ \bibinfo {author} {\bibfnamefont {J.}~\bibnamefont {Krause}},\ }in\
  \href@noop {} {\emph {\bibinfo {booktitle} {Super-Intense Laser-Atom
  Physics}}}\ (\bibinfo  {publisher} {Springer},\ \bibinfo {year} {1993})\ pp.\
  \bibinfo {pages} {95--110}\BibitemShut {NoStop}%
\bibitem [{\citenamefont {Corkum}(1993)}]{PhysRevLett.71.1994}%
  \BibitemOpen
  \bibfield  {author} {\bibinfo {author} {\bibfnamefont {P.~B.}\ \bibnamefont
  {Corkum}},\ }\href {\doibase 10.1103/PhysRevLett.71.1994} {\bibfield
  {journal} {\bibinfo  {journal} {Phys. Rev. Lett.}\ }\textbf {\bibinfo
  {volume} {71}},\ \bibinfo {pages} {1994} (\bibinfo {year}
  {1993})}\BibitemShut {NoStop}%
\bibitem [{\citenamefont {Lewenstein}\ \emph {et~al.}(1994)\citenamefont
  {Lewenstein}, \citenamefont {Balcou}, \citenamefont {Ivanov}, \citenamefont
  {L'Huillier},\ and\ \citenamefont {Corkum}}]{PhysRevA.49.2117}%
  \BibitemOpen
  \bibfield  {author} {\bibinfo {author} {\bibfnamefont {M.}~\bibnamefont
  {Lewenstein}}, \bibinfo {author} {\bibfnamefont {P.}~\bibnamefont {Balcou}},
  \bibinfo {author} {\bibfnamefont {M.~Y.}\ \bibnamefont {Ivanov}}, \bibinfo
  {author} {\bibfnamefont {A.}~\bibnamefont {L'Huillier}}, \ and\ \bibinfo
  {author} {\bibfnamefont {P.~B.}\ \bibnamefont {Corkum}},\ }\href {\doibase
  10.1103/PhysRevA.49.2117} {\bibfield  {journal} {\bibinfo  {journal} {Phys.
  Rev. A}\ }\textbf {\bibinfo {volume} {49}},\ \bibinfo {pages} {2117}
  (\bibinfo {year} {1994})}\BibitemShut {NoStop}%
\bibitem [{\citenamefont {Brabec}\ and\ \citenamefont
  {Krausz}(2000)}]{RevModPhys.72.545}%
  \BibitemOpen
  \bibfield  {author} {\bibinfo {author} {\bibfnamefont {T.}~\bibnamefont
  {Brabec}}\ and\ \bibinfo {author} {\bibfnamefont {F.}~\bibnamefont
  {Krausz}},\ }\href {\doibase 10.1103/RevModPhys.72.545} {\bibfield  {journal}
  {\bibinfo  {journal} {Rev. Mod. Phys.}\ }\textbf {\bibinfo {volume} {72}},\
  \bibinfo {pages} {545} (\bibinfo {year} {2000})}\BibitemShut {NoStop}%
\bibitem [{\citenamefont {Krausz}\ and\ \citenamefont
  {Ivanov}(2009)}]{RevModPhys.81.163}%
  \BibitemOpen
  \bibfield  {author} {\bibinfo {author} {\bibfnamefont {F.}~\bibnamefont
  {Krausz}}\ and\ \bibinfo {author} {\bibfnamefont {M.}~\bibnamefont
  {Ivanov}},\ }\href {\doibase 10.1103/RevModPhys.81.163} {\bibfield  {journal}
  {\bibinfo  {journal} {Rev. Mod. Phys.}\ }\textbf {\bibinfo {volume} {81}},\
  \bibinfo {pages} {163} (\bibinfo {year} {2009})}\BibitemShut {NoStop}%
\bibitem [{\citenamefont {Goulielmakis}\ \emph {et~al.}(2010)\citenamefont
  {Goulielmakis}, \citenamefont {Loh}, \citenamefont {Wirth}, \citenamefont
  {Santra}, \citenamefont {Rohringer}, \citenamefont {Yakovlev}, \citenamefont
  {Zherebtsov}, \citenamefont {Pfeifer}, \citenamefont {Azzeer}, \citenamefont
  {Kling}, \citenamefont {Leone},\ and\ \citenamefont
  {Krausz}}]{Goulielmakis2010}%
  \BibitemOpen
  \bibfield  {author} {\bibinfo {author} {\bibfnamefont {E.}~\bibnamefont
  {Goulielmakis}}, \bibinfo {author} {\bibfnamefont {Z.-H.}\ \bibnamefont
  {Loh}}, \bibinfo {author} {\bibfnamefont {A.}~\bibnamefont {Wirth}}, \bibinfo
  {author} {\bibfnamefont {R.}~\bibnamefont {Santra}}, \bibinfo {author}
  {\bibfnamefont {N.}~\bibnamefont {Rohringer}}, \bibinfo {author}
  {\bibfnamefont {V.~S.}\ \bibnamefont {Yakovlev}}, \bibinfo {author}
  {\bibfnamefont {S.}~\bibnamefont {Zherebtsov}}, \bibinfo {author}
  {\bibfnamefont {T.}~\bibnamefont {Pfeifer}}, \bibinfo {author} {\bibfnamefont
  {A.~M.}\ \bibnamefont {Azzeer}}, \bibinfo {author} {\bibfnamefont {M.~F.}\
  \bibnamefont {Kling}}, \bibinfo {author} {\bibfnamefont {S.~R.}\ \bibnamefont
  {Leone}}, \ and\ \bibinfo {author} {\bibfnamefont {F.}~\bibnamefont
  {Krausz}},\ }\href {\doibase 10.1038/nature09212} {\bibfield  {journal}
  {\bibinfo  {journal} {Nature}\ }\textbf {\bibinfo {volume} {466}},\ \bibinfo
  {pages} {739} (\bibinfo {year} {2010})}\BibitemShut {NoStop}%
\bibitem [{\citenamefont {Beck}\ \emph {et~al.}(2014)\citenamefont {Beck},
  \citenamefont {Bernhardt}, \citenamefont {Warrick}, \citenamefont {Wu},
  \citenamefont {Chen}, \citenamefont {Gaarde}, \citenamefont {Schafer},
  \citenamefont {Neumark},\ and\ \citenamefont {Leone}}]{Beck_2014}%
  \BibitemOpen
  \bibfield  {author} {\bibinfo {author} {\bibfnamefont {A.~R.}\ \bibnamefont
  {Beck}}, \bibinfo {author} {\bibfnamefont {B.}~\bibnamefont {Bernhardt}},
  \bibinfo {author} {\bibfnamefont {E.~R.}\ \bibnamefont {Warrick}}, \bibinfo
  {author} {\bibfnamefont {M.}~\bibnamefont {Wu}}, \bibinfo {author}
  {\bibfnamefont {S.}~\bibnamefont {Chen}}, \bibinfo {author} {\bibfnamefont
  {M.~B.}\ \bibnamefont {Gaarde}}, \bibinfo {author} {\bibfnamefont {K.~J.}\
  \bibnamefont {Schafer}}, \bibinfo {author} {\bibfnamefont {D.~M.}\
  \bibnamefont {Neumark}}, \ and\ \bibinfo {author} {\bibfnamefont {S.~R.}\
  \bibnamefont {Leone}},\ }\href {\doibase 10.1088/1367-2630/16/11/113016}
  {\bibfield  {journal} {\bibinfo  {journal} {New Journal of Physics}\ }\textbf
  {\bibinfo {volume} {16}},\ \bibinfo {pages} {113016} (\bibinfo {year}
  {2014})}\BibitemShut {NoStop}%
\bibitem [{\citenamefont {Schultze}\ \emph {et~al.}(2014)\citenamefont
  {Schultze}, \citenamefont {Ramasesha}, \citenamefont {Pemmaraju},
  \citenamefont {Sato}, \citenamefont {Whitmore}, \citenamefont {Gandman},
  \citenamefont {Prell}, \citenamefont {Borja}, \citenamefont {Prendergast},
  \citenamefont {Yabana}, \citenamefont {Neumark},\ and\ \citenamefont
  {Leone}}]{Schultze1348}%
  \BibitemOpen
  \bibfield  {author} {\bibinfo {author} {\bibfnamefont {M.}~\bibnamefont
  {Schultze}}, \bibinfo {author} {\bibfnamefont {K.}~\bibnamefont {Ramasesha}},
  \bibinfo {author} {\bibfnamefont {C.}~\bibnamefont {Pemmaraju}}, \bibinfo
  {author} {\bibfnamefont {S.}~\bibnamefont {Sato}}, \bibinfo {author}
  {\bibfnamefont {D.}~\bibnamefont {Whitmore}}, \bibinfo {author}
  {\bibfnamefont {A.}~\bibnamefont {Gandman}}, \bibinfo {author} {\bibfnamefont
  {J.~S.}\ \bibnamefont {Prell}}, \bibinfo {author} {\bibfnamefont {L.~J.}\
  \bibnamefont {Borja}}, \bibinfo {author} {\bibfnamefont {D.}~\bibnamefont
  {Prendergast}}, \bibinfo {author} {\bibfnamefont {K.}~\bibnamefont {Yabana}},
  \bibinfo {author} {\bibfnamefont {D.~M.}\ \bibnamefont {Neumark}}, \ and\
  \bibinfo {author} {\bibfnamefont {S.~R.}\ \bibnamefont {Leone}},\ }\href
  {\doibase 10.1126/science.1260311} {\bibfield  {journal} {\bibinfo  {journal}
  {Science}\ }\textbf {\bibinfo {volume} {346}},\ \bibinfo {pages} {1348}
  (\bibinfo {year} {2014})}\BibitemShut {NoStop}%
\bibitem [{\citenamefont {Lucchini}\ \emph {et~al.}(2016)\citenamefont
  {Lucchini}, \citenamefont {Sato}, \citenamefont {Ludwig}, \citenamefont
  {Herrmann}, \citenamefont {Volkov}, \citenamefont {Kasmi}, \citenamefont
  {Shinohara}, \citenamefont {Yabana}, \citenamefont {Gallmann},\ and\
  \citenamefont {Keller}}]{Lucchini916}%
  \BibitemOpen
  \bibfield  {author} {\bibinfo {author} {\bibfnamefont {M.}~\bibnamefont
  {Lucchini}}, \bibinfo {author} {\bibfnamefont {S.~A.}\ \bibnamefont {Sato}},
  \bibinfo {author} {\bibfnamefont {A.}~\bibnamefont {Ludwig}}, \bibinfo
  {author} {\bibfnamefont {J.}~\bibnamefont {Herrmann}}, \bibinfo {author}
  {\bibfnamefont {M.}~\bibnamefont {Volkov}}, \bibinfo {author} {\bibfnamefont
  {L.}~\bibnamefont {Kasmi}}, \bibinfo {author} {\bibfnamefont
  {Y.}~\bibnamefont {Shinohara}}, \bibinfo {author} {\bibfnamefont
  {K.}~\bibnamefont {Yabana}}, \bibinfo {author} {\bibfnamefont
  {L.}~\bibnamefont {Gallmann}}, \ and\ \bibinfo {author} {\bibfnamefont
  {U.}~\bibnamefont {Keller}},\ }\href {\doibase 10.1126/science.aag1268}
  {\bibfield  {journal} {\bibinfo  {journal} {Science}\ }\textbf {\bibinfo
  {volume} {353}},\ \bibinfo {pages} {916} (\bibinfo {year}
  {2016})}\BibitemShut {NoStop}%
\bibitem [{\citenamefont {Z{\"u}rch}\ \emph {et~al.}(2017)\citenamefont
  {Z{\"u}rch}, \citenamefont {Chang}, \citenamefont {Borja}, \citenamefont
  {Kraus}, \citenamefont {Cushing}, \citenamefont {Gandman}, \citenamefont
  {Kaplan}, \citenamefont {Oh}, \citenamefont {Prell}, \citenamefont
  {Prendergast}, \citenamefont {Pemmaraju}, \citenamefont {Neumark},\ and\
  \citenamefont {Leone}}]{Zurch2017}%
  \BibitemOpen
  \bibfield  {author} {\bibinfo {author} {\bibfnamefont {M.}~\bibnamefont
  {Z{\"u}rch}}, \bibinfo {author} {\bibfnamefont {H.-T.}\ \bibnamefont
  {Chang}}, \bibinfo {author} {\bibfnamefont {L.~J.}\ \bibnamefont {Borja}},
  \bibinfo {author} {\bibfnamefont {P.~M.}\ \bibnamefont {Kraus}}, \bibinfo
  {author} {\bibfnamefont {S.~K.}\ \bibnamefont {Cushing}}, \bibinfo {author}
  {\bibfnamefont {A.}~\bibnamefont {Gandman}}, \bibinfo {author} {\bibfnamefont
  {C.~J.}\ \bibnamefont {Kaplan}}, \bibinfo {author} {\bibfnamefont {M.~H.}\
  \bibnamefont {Oh}}, \bibinfo {author} {\bibfnamefont {J.~S.}\ \bibnamefont
  {Prell}}, \bibinfo {author} {\bibfnamefont {D.}~\bibnamefont {Prendergast}},
  \bibinfo {author} {\bibfnamefont {C.~D.}\ \bibnamefont {Pemmaraju}}, \bibinfo
  {author} {\bibfnamefont {D.~M.}\ \bibnamefont {Neumark}}, \ and\ \bibinfo
  {author} {\bibfnamefont {S.~R.}\ \bibnamefont {Leone}},\ }\href {\doibase
  10.1038/ncomms15734} {\bibfield  {journal} {\bibinfo  {journal} {Nature
  Communications}\ }\textbf {\bibinfo {volume} {8}},\ \bibinfo {pages} {15734}
  (\bibinfo {year} {2017})}\BibitemShut {NoStop}%
\bibitem [{\citenamefont {Siegrist}\ \emph {et~al.}(2019)\citenamefont
  {Siegrist}, \citenamefont {Gessner}, \citenamefont {Ossiander}, \citenamefont
  {Denker}, \citenamefont {Chang}, \citenamefont {Schr{\"o}der}, \citenamefont
  {Guggenmos}, \citenamefont {Cui}, \citenamefont {Walowski}, \citenamefont
  {Martens}, \citenamefont {Dewhurst}, \citenamefont {Kleineberg},
  \citenamefont {M{\"u}nzenberg}, \citenamefont {Sharma},\ and\ \citenamefont
  {Schultze}}]{Siegrist2019}%
  \BibitemOpen
  \bibfield  {author} {\bibinfo {author} {\bibfnamefont {F.}~\bibnamefont
  {Siegrist}}, \bibinfo {author} {\bibfnamefont {J.~A.}\ \bibnamefont
  {Gessner}}, \bibinfo {author} {\bibfnamefont {M.}~\bibnamefont {Ossiander}},
  \bibinfo {author} {\bibfnamefont {C.}~\bibnamefont {Denker}}, \bibinfo
  {author} {\bibfnamefont {Y.-P.}\ \bibnamefont {Chang}}, \bibinfo {author}
  {\bibfnamefont {M.~C.}\ \bibnamefont {Schr{\"o}der}}, \bibinfo {author}
  {\bibfnamefont {A.}~\bibnamefont {Guggenmos}}, \bibinfo {author}
  {\bibfnamefont {Y.}~\bibnamefont {Cui}}, \bibinfo {author} {\bibfnamefont
  {J.}~\bibnamefont {Walowski}}, \bibinfo {author} {\bibfnamefont
  {U.}~\bibnamefont {Martens}}, \bibinfo {author} {\bibfnamefont {J.~K.}\
  \bibnamefont {Dewhurst}}, \bibinfo {author} {\bibfnamefont {U.}~\bibnamefont
  {Kleineberg}}, \bibinfo {author} {\bibfnamefont {M.}~\bibnamefont
  {M{\"u}nzenberg}}, \bibinfo {author} {\bibfnamefont {S.}~\bibnamefont
  {Sharma}}, \ and\ \bibinfo {author} {\bibfnamefont {M.}~\bibnamefont
  {Schultze}},\ }\href {\doibase 10.1038/s41586-019-1333-x} {\bibfield
  {journal} {\bibinfo  {journal} {Nature}\ }\textbf {\bibinfo {volume} {571}},\
  \bibinfo {pages} {240} (\bibinfo {year} {2019})}\BibitemShut {NoStop}%
\bibitem [{\citenamefont {Volkov}\ \emph {et~al.}(2019)\citenamefont {Volkov},
  \citenamefont {Sato}, \citenamefont {Schlaepfer}, \citenamefont {Kasmi},
  \citenamefont {Hartmann}, \citenamefont {Lucchini}, \citenamefont {Gallmann},
  \citenamefont {Rubio},\ and\ \citenamefont {Keller}}]{Volkov2019}%
  \BibitemOpen
  \bibfield  {author} {\bibinfo {author} {\bibfnamefont {M.}~\bibnamefont
  {Volkov}}, \bibinfo {author} {\bibfnamefont {S.~A.}\ \bibnamefont {Sato}},
  \bibinfo {author} {\bibfnamefont {F.}~\bibnamefont {Schlaepfer}}, \bibinfo
  {author} {\bibfnamefont {L.}~\bibnamefont {Kasmi}}, \bibinfo {author}
  {\bibfnamefont {N.}~\bibnamefont {Hartmann}}, \bibinfo {author}
  {\bibfnamefont {M.}~\bibnamefont {Lucchini}}, \bibinfo {author}
  {\bibfnamefont {L.}~\bibnamefont {Gallmann}}, \bibinfo {author}
  {\bibfnamefont {A.}~\bibnamefont {Rubio}}, \ and\ \bibinfo {author}
  {\bibfnamefont {U.}~\bibnamefont {Keller}},\ }\href {\doibase
  10.1038/s41567-019-0602-9} {\bibfield  {journal} {\bibinfo  {journal} {Nature
  Physics}\ }\textbf {\bibinfo {volume} {15}},\ \bibinfo {pages} {1145}
  (\bibinfo {year} {2019})}\BibitemShut {NoStop}%
\bibitem [{\citenamefont {Ghimire}\ \emph {et~al.}(2011)\citenamefont
  {Ghimire}, \citenamefont {DiChiara}, \citenamefont {Sistrunk}, \citenamefont
  {Agostini}, \citenamefont {DiMauro},\ and\ \citenamefont
  {Reis}}]{Ghimire2011}%
  \BibitemOpen
  \bibfield  {author} {\bibinfo {author} {\bibfnamefont {S.}~\bibnamefont
  {Ghimire}}, \bibinfo {author} {\bibfnamefont {A.~D.}\ \bibnamefont
  {DiChiara}}, \bibinfo {author} {\bibfnamefont {E.}~\bibnamefont {Sistrunk}},
  \bibinfo {author} {\bibfnamefont {P.}~\bibnamefont {Agostini}}, \bibinfo
  {author} {\bibfnamefont {L.~F.}\ \bibnamefont {DiMauro}}, \ and\ \bibinfo
  {author} {\bibfnamefont {D.~A.}\ \bibnamefont {Reis}},\ }\href {\doibase
  10.1038/nphys1847} {\bibfield  {journal} {\bibinfo  {journal} {Nature
  Physics}\ }\textbf {\bibinfo {volume} {7}},\ \bibinfo {pages} {138} (\bibinfo
  {year} {2011})}\BibitemShut {NoStop}%
\bibitem [{\citenamefont {Ghimire}\ and\ \citenamefont
  {Reis}(2019)}]{Ghimire2019}%
  \BibitemOpen
  \bibfield  {author} {\bibinfo {author} {\bibfnamefont {S.}~\bibnamefont
  {Ghimire}}\ and\ \bibinfo {author} {\bibfnamefont {D.~A.}\ \bibnamefont
  {Reis}},\ }\href {\doibase 10.1038/s41567-018-0315-5} {\bibfield  {journal}
  {\bibinfo  {journal} {Nature Physics}\ }\textbf {\bibinfo {volume} {15}},\
  \bibinfo {pages} {10} (\bibinfo {year} {2019})}\BibitemShut {NoStop}%
\bibitem [{\citenamefont {Vampa}\ \emph
  {et~al.}(2015{\natexlab{a}})\citenamefont {Vampa}, \citenamefont {Hammond},
  \citenamefont {Thir\'e}, \citenamefont {Schmidt}, \citenamefont {L\'egar\'e},
  \citenamefont {McDonald}, \citenamefont {Brabec}, \citenamefont {Klug},\ and\
  \citenamefont {Corkum}}]{PhysRevLett.115.193603}%
  \BibitemOpen
  \bibfield  {author} {\bibinfo {author} {\bibfnamefont {G.}~\bibnamefont
  {Vampa}}, \bibinfo {author} {\bibfnamefont {T.~J.}\ \bibnamefont {Hammond}},
  \bibinfo {author} {\bibfnamefont {N.}~\bibnamefont {Thir\'e}}, \bibinfo
  {author} {\bibfnamefont {B.~E.}\ \bibnamefont {Schmidt}}, \bibinfo {author}
  {\bibfnamefont {F.}~\bibnamefont {L\'egar\'e}}, \bibinfo {author}
  {\bibfnamefont {C.~R.}\ \bibnamefont {McDonald}}, \bibinfo {author}
  {\bibfnamefont {T.}~\bibnamefont {Brabec}}, \bibinfo {author} {\bibfnamefont
  {D.~D.}\ \bibnamefont {Klug}}, \ and\ \bibinfo {author} {\bibfnamefont
  {P.~B.}\ \bibnamefont {Corkum}},\ }\href {\doibase
  10.1103/PhysRevLett.115.193603} {\bibfield  {journal} {\bibinfo  {journal}
  {Phys. Rev. Lett.}\ }\textbf {\bibinfo {volume} {115}},\ \bibinfo {pages}
  {193603} (\bibinfo {year} {2015}{\natexlab{a}})}\BibitemShut {NoStop}%
\bibitem [{\citenamefont {Chac{\'o}n}\ \emph {et~al.}(2018)\citenamefont
  {Chac{\'o}n}, \citenamefont {Zhu}, \citenamefont {Kelly}, \citenamefont
  {Dauphin}, \citenamefont {Pisanty}, \citenamefont {Pic{\'o}n}, \citenamefont
  {Ticknor}, \citenamefont {Ciappina}, \citenamefont {Saxena},\ and\
  \citenamefont {Lewenstein}}]{chacon2018observing}%
  \BibitemOpen
  \bibfield  {author} {\bibinfo {author} {\bibfnamefont {A.}~\bibnamefont
  {Chac{\'o}n}}, \bibinfo {author} {\bibfnamefont {W.}~\bibnamefont {Zhu}},
  \bibinfo {author} {\bibfnamefont {S.~P.}\ \bibnamefont {Kelly}}, \bibinfo
  {author} {\bibfnamefont {A.}~\bibnamefont {Dauphin}}, \bibinfo {author}
  {\bibfnamefont {E.}~\bibnamefont {Pisanty}}, \bibinfo {author} {\bibfnamefont
  {A.}~\bibnamefont {Pic{\'o}n}}, \bibinfo {author} {\bibfnamefont
  {C.}~\bibnamefont {Ticknor}}, \bibinfo {author} {\bibfnamefont {M.~F.}\
  \bibnamefont {Ciappina}}, \bibinfo {author} {\bibfnamefont {A.}~\bibnamefont
  {Saxena}}, \ and\ \bibinfo {author} {\bibfnamefont {M.}~\bibnamefont
  {Lewenstein}},\ }\href@noop {} {\bibfield  {journal} {\bibinfo  {journal}
  {arXiv:1807.01616 [cond-mat.mes-hall]}\ } (\bibinfo {year}
  {2018})}\BibitemShut {NoStop}%
\bibitem [{\citenamefont {Silva}\ \emph {et~al.}(2019)\citenamefont {Silva},
  \citenamefont {Jim{\'e}nez-Gal{\'a}n}, \citenamefont {Amorim}, \citenamefont
  {Smirnova},\ and\ \citenamefont {Ivanov}}]{Silva2019}%
  \BibitemOpen
  \bibfield  {author} {\bibinfo {author} {\bibfnamefont {R.~E.~F.}\
  \bibnamefont {Silva}}, \bibinfo {author} {\bibfnamefont {{\'A}.}~\bibnamefont
  {Jim{\'e}nez-Gal{\'a}n}}, \bibinfo {author} {\bibfnamefont {B.}~\bibnamefont
  {Amorim}}, \bibinfo {author} {\bibfnamefont {O.}~\bibnamefont {Smirnova}}, \
  and\ \bibinfo {author} {\bibfnamefont {M.}~\bibnamefont {Ivanov}},\ }\href
  {\doibase 10.1038/s41566-019-0516-1} {\bibfield  {journal} {\bibinfo
  {journal} {Nature Photonics}\ }\textbf {\bibinfo {volume} {13}},\ \bibinfo
  {pages} {849} (\bibinfo {year} {2019})}\BibitemShut {NoStop}%
\bibitem [{\citenamefont {Schubert}\ \emph {et~al.}(2014)\citenamefont
  {Schubert}, \citenamefont {Hohenleutner}, \citenamefont {Langer},
  \citenamefont {Urbanek}, \citenamefont {Lange}, \citenamefont {Huttner},
  \citenamefont {Golde}, \citenamefont {Meier}, \citenamefont {Kira},
  \citenamefont {Koch},\ and\ \citenamefont {Huber}}]{Schubert2014}%
  \BibitemOpen
  \bibfield  {author} {\bibinfo {author} {\bibfnamefont {O.}~\bibnamefont
  {Schubert}}, \bibinfo {author} {\bibfnamefont {M.}~\bibnamefont
  {Hohenleutner}}, \bibinfo {author} {\bibfnamefont {F.}~\bibnamefont
  {Langer}}, \bibinfo {author} {\bibfnamefont {B.}~\bibnamefont {Urbanek}},
  \bibinfo {author} {\bibfnamefont {C.}~\bibnamefont {Lange}}, \bibinfo
  {author} {\bibfnamefont {U.}~\bibnamefont {Huttner}}, \bibinfo {author}
  {\bibfnamefont {D.}~\bibnamefont {Golde}}, \bibinfo {author} {\bibfnamefont
  {T.}~\bibnamefont {Meier}}, \bibinfo {author} {\bibfnamefont
  {M.}~\bibnamefont {Kira}}, \bibinfo {author} {\bibfnamefont {S.~W.}\
  \bibnamefont {Koch}}, \ and\ \bibinfo {author} {\bibfnamefont
  {R.}~\bibnamefont {Huber}},\ }\href {\doibase 10.1038/nphoton.2013.349}
  {\bibfield  {journal} {\bibinfo  {journal} {Nature Photonics}\ }\textbf
  {\bibinfo {volume} {8}},\ \bibinfo {pages} {119} (\bibinfo {year}
  {2014})}\BibitemShut {NoStop}%
\bibitem [{\citenamefont {Luu}\ \emph {et~al.}(2015)\citenamefont {Luu},
  \citenamefont {Garg}, \citenamefont {Kruchinin}, \citenamefont {Moulet},
  \citenamefont {Hassan},\ and\ \citenamefont {Goulielmakis}}]{Luu2015}%
  \BibitemOpen
  \bibfield  {author} {\bibinfo {author} {\bibfnamefont {T.~T.}\ \bibnamefont
  {Luu}}, \bibinfo {author} {\bibfnamefont {M.}~\bibnamefont {Garg}}, \bibinfo
  {author} {\bibfnamefont {S.~Y.}\ \bibnamefont {Kruchinin}}, \bibinfo {author}
  {\bibfnamefont {A.}~\bibnamefont {Moulet}}, \bibinfo {author} {\bibfnamefont
  {M.~T.}\ \bibnamefont {Hassan}}, \ and\ \bibinfo {author} {\bibfnamefont
  {E.}~\bibnamefont {Goulielmakis}},\ }\href {\doibase 10.1038/nature14456}
  {\bibfield  {journal} {\bibinfo  {journal} {Nature}\ }\textbf {\bibinfo
  {volume} {521}},\ \bibinfo {pages} {498} (\bibinfo {year}
  {2015})}\BibitemShut {NoStop}%
\bibitem [{\citenamefont {You}\ \emph {et~al.}(2017)\citenamefont {You},
  \citenamefont {Reis},\ and\ \citenamefont {Ghimire}}]{You2017}%
  \BibitemOpen
  \bibfield  {author} {\bibinfo {author} {\bibfnamefont {Y.~S.}\ \bibnamefont
  {You}}, \bibinfo {author} {\bibfnamefont {D.}~\bibnamefont {Reis}}, \ and\
  \bibinfo {author} {\bibfnamefont {S.}~\bibnamefont {Ghimire}},\ }\href
  {\doibase 10.1038/nphys3955} {\bibfield  {journal} {\bibinfo  {journal}
  {Nature Physics}\ }\textbf {\bibinfo {volume} {13}},\ \bibinfo {pages} {345}
  (\bibinfo {year} {2017})}\BibitemShut {NoStop}%
\bibitem [{\citenamefont {Yoshikawa}\ \emph {et~al.}(2017)\citenamefont
  {Yoshikawa}, \citenamefont {Tamaya},\ and\ \citenamefont
  {Tanaka}}]{Yoshikawa736}%
  \BibitemOpen
  \bibfield  {author} {\bibinfo {author} {\bibfnamefont {N.}~\bibnamefont
  {Yoshikawa}}, \bibinfo {author} {\bibfnamefont {T.}~\bibnamefont {Tamaya}}, \
  and\ \bibinfo {author} {\bibfnamefont {K.}~\bibnamefont {Tanaka}},\ }\href
  {\doibase 10.1126/science.aam8861} {\bibfield  {journal} {\bibinfo  {journal}
  {Science}\ }\textbf {\bibinfo {volume} {356}},\ \bibinfo {pages} {736}
  (\bibinfo {year} {2017})}\BibitemShut {NoStop}%
\bibitem [{\citenamefont {Sanari}\ \emph {et~al.}(2020)\citenamefont {Sanari},
  \citenamefont {Hirori}, \citenamefont {Aharen}, \citenamefont {Tahara},
  \citenamefont {Shinohara}, \citenamefont {Ishikawa}, \citenamefont {Otobe},
  \citenamefont {Xia}, \citenamefont {Ishii}, \citenamefont {Itatani},
  \citenamefont {Sato},\ and\ \citenamefont {Kanemitsu}}]{PhysRevB.102.041125}%
  \BibitemOpen
  \bibfield  {author} {\bibinfo {author} {\bibfnamefont {Y.}~\bibnamefont
  {Sanari}}, \bibinfo {author} {\bibfnamefont {H.}~\bibnamefont {Hirori}},
  \bibinfo {author} {\bibfnamefont {T.}~\bibnamefont {Aharen}}, \bibinfo
  {author} {\bibfnamefont {H.}~\bibnamefont {Tahara}}, \bibinfo {author}
  {\bibfnamefont {Y.}~\bibnamefont {Shinohara}}, \bibinfo {author}
  {\bibfnamefont {K.~L.}\ \bibnamefont {Ishikawa}}, \bibinfo {author}
  {\bibfnamefont {T.}~\bibnamefont {Otobe}}, \bibinfo {author} {\bibfnamefont
  {P.}~\bibnamefont {Xia}}, \bibinfo {author} {\bibfnamefont {N.}~\bibnamefont
  {Ishii}}, \bibinfo {author} {\bibfnamefont {J.}~\bibnamefont {Itatani}},
  \bibinfo {author} {\bibfnamefont {S.~A.}\ \bibnamefont {Sato}}, \ and\
  \bibinfo {author} {\bibfnamefont {Y.}~\bibnamefont {Kanemitsu}},\ }\href
  {\doibase 10.1103/PhysRevB.102.041125} {\bibfield  {journal} {\bibinfo
  {journal} {Phys. Rev. B}\ }\textbf {\bibinfo {volume} {102}},\ \bibinfo
  {pages} {041125} (\bibinfo {year} {2020})}\BibitemShut {NoStop}%
\bibitem [{\citenamefont {Golde}\ \emph {et~al.}(2008)\citenamefont {Golde},
  \citenamefont {Meier},\ and\ \citenamefont {Koch}}]{PhysRevB.77.075330}%
  \BibitemOpen
  \bibfield  {author} {\bibinfo {author} {\bibfnamefont {D.}~\bibnamefont
  {Golde}}, \bibinfo {author} {\bibfnamefont {T.}~\bibnamefont {Meier}}, \ and\
  \bibinfo {author} {\bibfnamefont {S.~W.}\ \bibnamefont {Koch}},\ }\href
  {\doibase 10.1103/PhysRevB.77.075330} {\bibfield  {journal} {\bibinfo
  {journal} {Phys. Rev. B}\ }\textbf {\bibinfo {volume} {77}},\ \bibinfo
  {pages} {075330} (\bibinfo {year} {2008})}\BibitemShut {NoStop}%
\bibitem [{\citenamefont {Vampa}\ \emph {et~al.}(2014)\citenamefont {Vampa},
  \citenamefont {McDonald}, \citenamefont {Orlando}, \citenamefont {Klug},
  \citenamefont {Corkum},\ and\ \citenamefont
  {Brabec}}]{PhysRevLett.113.073901}%
  \BibitemOpen
  \bibfield  {author} {\bibinfo {author} {\bibfnamefont {G.}~\bibnamefont
  {Vampa}}, \bibinfo {author} {\bibfnamefont {C.~R.}\ \bibnamefont {McDonald}},
  \bibinfo {author} {\bibfnamefont {G.}~\bibnamefont {Orlando}}, \bibinfo
  {author} {\bibfnamefont {D.~D.}\ \bibnamefont {Klug}}, \bibinfo {author}
  {\bibfnamefont {P.~B.}\ \bibnamefont {Corkum}}, \ and\ \bibinfo {author}
  {\bibfnamefont {T.}~\bibnamefont {Brabec}},\ }\href {\doibase
  10.1103/PhysRevLett.113.073901} {\bibfield  {journal} {\bibinfo  {journal}
  {Phys. Rev. Lett.}\ }\textbf {\bibinfo {volume} {113}},\ \bibinfo {pages}
  {073901} (\bibinfo {year} {2014})}\BibitemShut {NoStop}%
\bibitem [{\citenamefont {Vampa}\ \emph
  {et~al.}(2015{\natexlab{b}})\citenamefont {Vampa}, \citenamefont {McDonald},
  \citenamefont {Orlando}, \citenamefont {Corkum},\ and\ \citenamefont
  {Brabec}}]{PhysRevB.91.064302}%
  \BibitemOpen
  \bibfield  {author} {\bibinfo {author} {\bibfnamefont {G.}~\bibnamefont
  {Vampa}}, \bibinfo {author} {\bibfnamefont {C.~R.}\ \bibnamefont {McDonald}},
  \bibinfo {author} {\bibfnamefont {G.}~\bibnamefont {Orlando}}, \bibinfo
  {author} {\bibfnamefont {P.~B.}\ \bibnamefont {Corkum}}, \ and\ \bibinfo
  {author} {\bibfnamefont {T.}~\bibnamefont {Brabec}},\ }\href {\doibase
  10.1103/PhysRevB.91.064302} {\bibfield  {journal} {\bibinfo  {journal} {Phys.
  Rev. B}\ }\textbf {\bibinfo {volume} {91}},\ \bibinfo {pages} {064302}
  (\bibinfo {year} {2015}{\natexlab{b}})}\BibitemShut {NoStop}%
\bibitem [{\citenamefont {Floss}\ \emph {et~al.}(2018)\citenamefont {Floss},
  \citenamefont {Lemell}, \citenamefont {Wachter}, \citenamefont {Smejkal},
  \citenamefont {Sato}, \citenamefont {Tong}, \citenamefont {Yabana},\ and\
  \citenamefont {Burgd\"orfer}}]{PhysRevA.97.011401}%
  \BibitemOpen
  \bibfield  {author} {\bibinfo {author} {\bibfnamefont {I.}~\bibnamefont
  {Floss}}, \bibinfo {author} {\bibfnamefont {C.}~\bibnamefont {Lemell}},
  \bibinfo {author} {\bibfnamefont {G.}~\bibnamefont {Wachter}}, \bibinfo
  {author} {\bibfnamefont {V.}~\bibnamefont {Smejkal}}, \bibinfo {author}
  {\bibfnamefont {S.~A.}\ \bibnamefont {Sato}}, \bibinfo {author}
  {\bibfnamefont {X.-M.}\ \bibnamefont {Tong}}, \bibinfo {author}
  {\bibfnamefont {K.}~\bibnamefont {Yabana}}, \ and\ \bibinfo {author}
  {\bibfnamefont {J.}~\bibnamefont {Burgd\"orfer}},\ }\href {\doibase
  10.1103/PhysRevA.97.011401} {\bibfield  {journal} {\bibinfo  {journal} {Phys.
  Rev. A}\ }\textbf {\bibinfo {volume} {97}},\ \bibinfo {pages} {011401}
  (\bibinfo {year} {2018})}\BibitemShut {NoStop}%
\bibitem [{\citenamefont {Ikemachi}\ \emph {et~al.}(2017)\citenamefont
  {Ikemachi}, \citenamefont {Shinohara}, \citenamefont {Sato}, \citenamefont
  {Yumoto}, \citenamefont {Kuwata-Gonokami},\ and\ \citenamefont
  {Ishikawa}}]{PhysRevA.95.043416}%
  \BibitemOpen
  \bibfield  {author} {\bibinfo {author} {\bibfnamefont {T.}~\bibnamefont
  {Ikemachi}}, \bibinfo {author} {\bibfnamefont {Y.}~\bibnamefont {Shinohara}},
  \bibinfo {author} {\bibfnamefont {T.}~\bibnamefont {Sato}}, \bibinfo {author}
  {\bibfnamefont {J.}~\bibnamefont {Yumoto}}, \bibinfo {author} {\bibfnamefont
  {M.}~\bibnamefont {Kuwata-Gonokami}}, \ and\ \bibinfo {author} {\bibfnamefont
  {K.~L.}\ \bibnamefont {Ishikawa}},\ }\href {\doibase
  10.1103/PhysRevA.95.043416} {\bibfield  {journal} {\bibinfo  {journal} {Phys.
  Rev. A}\ }\textbf {\bibinfo {volume} {95}},\ \bibinfo {pages} {043416}
  (\bibinfo {year} {2017})}\BibitemShut {NoStop}%
\bibitem [{\citenamefont {Tancogne-Dejean}\ \emph {et~al.}(2017)\citenamefont
  {Tancogne-Dejean}, \citenamefont {M\"ucke}, \citenamefont {K\"artner},\ and\
  \citenamefont {Rubio}}]{PhysRevLett.118.087403}%
  \BibitemOpen
  \bibfield  {author} {\bibinfo {author} {\bibfnamefont {N.}~\bibnamefont
  {Tancogne-Dejean}}, \bibinfo {author} {\bibfnamefont {O.~D.}\ \bibnamefont
  {M\"ucke}}, \bibinfo {author} {\bibfnamefont {F.~X.}\ \bibnamefont
  {K\"artner}}, \ and\ \bibinfo {author} {\bibfnamefont {A.}~\bibnamefont
  {Rubio}},\ }\href {\doibase 10.1103/PhysRevLett.118.087403} {\bibfield
  {journal} {\bibinfo  {journal} {Phys. Rev. Lett.}\ }\textbf {\bibinfo
  {volume} {118}},\ \bibinfo {pages} {087403} (\bibinfo {year}
  {2017})}\BibitemShut {NoStop}%
\bibitem [{\citenamefont {Schlaepfer}\ \emph {et~al.}(2018)\citenamefont
  {Schlaepfer}, \citenamefont {Lucchini}, \citenamefont {Sato}, \citenamefont
  {Volkov}, \citenamefont {Kasmi}, \citenamefont {Hartmann}, \citenamefont
  {Rubio}, \citenamefont {Gallmann},\ and\ \citenamefont
  {Keller}}]{Schlaepfer2018}%
  \BibitemOpen
  \bibfield  {author} {\bibinfo {author} {\bibfnamefont {F.}~\bibnamefont
  {Schlaepfer}}, \bibinfo {author} {\bibfnamefont {M.}~\bibnamefont
  {Lucchini}}, \bibinfo {author} {\bibfnamefont {S.~A.}\ \bibnamefont {Sato}},
  \bibinfo {author} {\bibfnamefont {M.}~\bibnamefont {Volkov}}, \bibinfo
  {author} {\bibfnamefont {L.}~\bibnamefont {Kasmi}}, \bibinfo {author}
  {\bibfnamefont {N.}~\bibnamefont {Hartmann}}, \bibinfo {author}
  {\bibfnamefont {A.}~\bibnamefont {Rubio}}, \bibinfo {author} {\bibfnamefont
  {L.}~\bibnamefont {Gallmann}}, \ and\ \bibinfo {author} {\bibfnamefont
  {U.}~\bibnamefont {Keller}},\ }\href {\doibase 10.1038/s41567-018-0069-0}
  {\bibfield  {journal} {\bibinfo  {journal} {Nature Physics}\ }\textbf
  {\bibinfo {volume} {14}},\ \bibinfo {pages} {560} (\bibinfo {year}
  {2018})}\BibitemShut {NoStop}%
\bibitem [{\citenamefont {Buades}\ \emph {et~al.}(2018)\citenamefont {Buades},
  \citenamefont {Pic{\'o}n}, \citenamefont {Le{\'o}n}, \citenamefont {Di~Palo},
  \citenamefont {Cousin}, \citenamefont {Cocchi}, \citenamefont {Pellegrin},
  \citenamefont {Martin}, \citenamefont {Ma{\~n}as-Valero}, \citenamefont
  {Coronado} \emph {et~al.}}]{buades2018attosecond}%
  \BibitemOpen
  \bibfield  {author} {\bibinfo {author} {\bibfnamefont {B.}~\bibnamefont
  {Buades}}, \bibinfo {author} {\bibfnamefont {A.}~\bibnamefont {Pic{\'o}n}},
  \bibinfo {author} {\bibfnamefont {I.}~\bibnamefont {Le{\'o}n}}, \bibinfo
  {author} {\bibfnamefont {N.}~\bibnamefont {Di~Palo}}, \bibinfo {author}
  {\bibfnamefont {S.~L.}\ \bibnamefont {Cousin}}, \bibinfo {author}
  {\bibfnamefont {C.}~\bibnamefont {Cocchi}}, \bibinfo {author} {\bibfnamefont
  {E.}~\bibnamefont {Pellegrin}}, \bibinfo {author} {\bibfnamefont {J.~H.}\
  \bibnamefont {Martin}}, \bibinfo {author} {\bibfnamefont {S.}~\bibnamefont
  {Ma{\~n}as-Valero}}, \bibinfo {author} {\bibfnamefont {E.}~\bibnamefont
  {Coronado}},  \emph {et~al.},\ }\href@noop {} {\bibfield  {journal} {\bibinfo
   {journal} {arXiv:1808.06493 [cond-mat.mtrl-sci]}\ } (\bibinfo {year}
  {2018})}\BibitemShut {NoStop}%
\bibitem [{\citenamefont {Golde}\ \emph {et~al.}(2009)\citenamefont {Golde},
  \citenamefont {Meier},\ and\ \citenamefont {Koch}}]{golde2009microscopic}%
  \BibitemOpen
  \bibfield  {author} {\bibinfo {author} {\bibfnamefont {D.}~\bibnamefont
  {Golde}}, \bibinfo {author} {\bibfnamefont {T.}~\bibnamefont {Meier}}, \ and\
  \bibinfo {author} {\bibfnamefont {S.~W.}\ \bibnamefont {Koch}},\ }\href@noop
  {} {\bibfield  {journal} {\bibinfo  {journal} {physica status solidi c}\
  }\textbf {\bibinfo {volume} {6}},\ \bibinfo {pages} {420} (\bibinfo {year}
  {2009})}\BibitemShut {NoStop}%
\bibitem [{\citenamefont {Budil}\ \emph {et~al.}(1993)\citenamefont {Budil},
  \citenamefont {Sali\`eres}, \citenamefont {L'Huillier}, \citenamefont
  {Ditmire},\ and\ \citenamefont {Perry}}]{PhysRevA.48.R3437}%
  \BibitemOpen
  \bibfield  {author} {\bibinfo {author} {\bibfnamefont {K.~S.}\ \bibnamefont
  {Budil}}, \bibinfo {author} {\bibfnamefont {P.}~\bibnamefont {Sali\`eres}},
  \bibinfo {author} {\bibfnamefont {A.}~\bibnamefont {L'Huillier}}, \bibinfo
  {author} {\bibfnamefont {T.}~\bibnamefont {Ditmire}}, \ and\ \bibinfo
  {author} {\bibfnamefont {M.~D.}\ \bibnamefont {Perry}},\ }\href {\doibase
  10.1103/PhysRevA.48.R3437} {\bibfield  {journal} {\bibinfo  {journal} {Phys.
  Rev. A}\ }\textbf {\bibinfo {volume} {48}},\ \bibinfo {pages} {R3437}
  (\bibinfo {year} {1993})}\BibitemShut {NoStop}%
\bibitem [{\citenamefont {Dietrich}\ \emph {et~al.}(1994)\citenamefont
  {Dietrich}, \citenamefont {Burnett}, \citenamefont {Ivanov},\ and\
  \citenamefont {Corkum}}]{PhysRevA.50.R3585}%
  \BibitemOpen
  \bibfield  {author} {\bibinfo {author} {\bibfnamefont {P.}~\bibnamefont
  {Dietrich}}, \bibinfo {author} {\bibfnamefont {N.~H.}\ \bibnamefont
  {Burnett}}, \bibinfo {author} {\bibfnamefont {M.}~\bibnamefont {Ivanov}}, \
  and\ \bibinfo {author} {\bibfnamefont {P.~B.}\ \bibnamefont {Corkum}},\
  }\href {\doibase 10.1103/PhysRevA.50.R3585} {\bibfield  {journal} {\bibinfo
  {journal} {Phys. Rev. A}\ }\textbf {\bibinfo {volume} {50}},\ \bibinfo
  {pages} {R3585} (\bibinfo {year} {1994})}\BibitemShut {NoStop}%
\bibitem [{\citenamefont {Liang}\ \emph {et~al.}(1994)\citenamefont {Liang},
  \citenamefont {Ammosov},\ and\ \citenamefont {Chin}}]{Liang_1994}%
  \BibitemOpen
  \bibfield  {author} {\bibinfo {author} {\bibfnamefont {Y.}~\bibnamefont
  {Liang}}, \bibinfo {author} {\bibfnamefont {M.~V.}\ \bibnamefont {Ammosov}},
  \ and\ \bibinfo {author} {\bibfnamefont {S.~L.}\ \bibnamefont {Chin}},\
  }\href {\doibase 10.1088/0953-4075/27/6/024} {\bibfield  {journal} {\bibinfo
  {journal} {Journal of Physics B: Atomic, Molecular and Optical Physics}\
  }\textbf {\bibinfo {volume} {27}},\ \bibinfo {pages} {1269} (\bibinfo {year}
  {1994})}\BibitemShut {NoStop}%
\bibitem [{\citenamefont {Sato}\ \emph
  {et~al.}(2019{\natexlab{a}})\citenamefont {Sato}, \citenamefont {McIver},
  \citenamefont {Nuske}, \citenamefont {Tang}, \citenamefont {Jotzu},
  \citenamefont {Schulte}, \citenamefont {H\"ubener}, \citenamefont
  {De~Giovannini}, \citenamefont {Mathey}, \citenamefont {Sentef},
  \citenamefont {Cavalleri},\ and\ \citenamefont {Rubio}}]{PhysRevB.99.214302}%
  \BibitemOpen
  \bibfield  {author} {\bibinfo {author} {\bibfnamefont {S.~A.}\ \bibnamefont
  {Sato}}, \bibinfo {author} {\bibfnamefont {J.~W.}\ \bibnamefont {McIver}},
  \bibinfo {author} {\bibfnamefont {M.}~\bibnamefont {Nuske}}, \bibinfo
  {author} {\bibfnamefont {P.}~\bibnamefont {Tang}}, \bibinfo {author}
  {\bibfnamefont {G.}~\bibnamefont {Jotzu}}, \bibinfo {author} {\bibfnamefont
  {B.}~\bibnamefont {Schulte}}, \bibinfo {author} {\bibfnamefont
  {H.}~\bibnamefont {H\"ubener}}, \bibinfo {author} {\bibfnamefont
  {U.}~\bibnamefont {De~Giovannini}}, \bibinfo {author} {\bibfnamefont
  {L.}~\bibnamefont {Mathey}}, \bibinfo {author} {\bibfnamefont {M.~A.}\
  \bibnamefont {Sentef}}, \bibinfo {author} {\bibfnamefont {A.}~\bibnamefont
  {Cavalleri}}, \ and\ \bibinfo {author} {\bibfnamefont {A.}~\bibnamefont
  {Rubio}},\ }\href {\doibase 10.1103/PhysRevB.99.214302} {\bibfield  {journal}
  {\bibinfo  {journal} {Phys. Rev. B}\ }\textbf {\bibinfo {volume} {99}},\
  \bibinfo {pages} {214302} (\bibinfo {year} {2019}{\natexlab{a}})}\BibitemShut
  {NoStop}%
\bibitem [{\citenamefont {Sato}\ \emph
  {et~al.}(2019{\natexlab{b}})\citenamefont {Sato}, \citenamefont {Tang},
  \citenamefont {Sentef}, \citenamefont {Giovannini}, \citenamefont
  {H\"ubener},\ and\ \citenamefont {Rubio}}]{sato_njp_2019}%
  \BibitemOpen
  \bibfield  {author} {\bibinfo {author} {\bibfnamefont {S.~A.}\ \bibnamefont
  {Sato}}, \bibinfo {author} {\bibfnamefont {P.}~\bibnamefont {Tang}}, \bibinfo
  {author} {\bibfnamefont {M.~A.}\ \bibnamefont {Sentef}}, \bibinfo {author}
  {\bibfnamefont {U.~D.}\ \bibnamefont {Giovannini}}, \bibinfo {author}
  {\bibfnamefont {H.}~\bibnamefont {H\"ubener}}, \ and\ \bibinfo {author}
  {\bibfnamefont {A.}~\bibnamefont {Rubio}},\ }\href {\doibase
  10.1088/1367-2630/ab3acf} {\bibfield  {journal} {\bibinfo  {journal} {New
  Journal of Physics}\ }\textbf {\bibinfo {volume} {21}},\ \bibinfo {pages}
  {093005} (\bibinfo {year} {2019}{\natexlab{b}})}\BibitemShut {NoStop}%
\bibitem [{\citenamefont {Trevisanutto}\ \emph {et~al.}(2008)\citenamefont
  {Trevisanutto}, \citenamefont {Giorgetti}, \citenamefont {Reining},
  \citenamefont {Ladisa},\ and\ \citenamefont
  {Olevano}}]{PhysRevLett.101.226405}%
  \BibitemOpen
  \bibfield  {author} {\bibinfo {author} {\bibfnamefont {P.~E.}\ \bibnamefont
  {Trevisanutto}}, \bibinfo {author} {\bibfnamefont {C.}~\bibnamefont
  {Giorgetti}}, \bibinfo {author} {\bibfnamefont {L.}~\bibnamefont {Reining}},
  \bibinfo {author} {\bibfnamefont {M.}~\bibnamefont {Ladisa}}, \ and\ \bibinfo
  {author} {\bibfnamefont {V.}~\bibnamefont {Olevano}},\ }\href {\doibase
  10.1103/PhysRevLett.101.226405} {\bibfield  {journal} {\bibinfo  {journal}
  {Phys. Rev. Lett.}\ }\textbf {\bibinfo {volume} {101}},\ \bibinfo {pages}
  {226405} (\bibinfo {year} {2008})}\BibitemShut {NoStop}%
\bibitem [{Sup()}]{SuppMat}%
  \BibitemOpen
  \href@noop {} {\bibinfo  {journal} {See Supplemental Material for more
  details}\ }\BibitemShut {NoStop}%
\bibitem [{\citenamefont {Sato}\ \emph {et~al.}(2018)\citenamefont {Sato},
  \citenamefont {Lucchini}, \citenamefont {Volkov}, \citenamefont {Schlaepfer},
  \citenamefont {Gallmann}, \citenamefont {Keller},\ and\ \citenamefont
  {Rubio}}]{PhysRevB.98.035202}%
  \BibitemOpen
\bibfield  {journal} {  }\bibfield  {author} {\bibinfo {author} {\bibfnamefont
  {S.~A.}\ \bibnamefont {Sato}}, \bibinfo {author} {\bibfnamefont
  {M.}~\bibnamefont {Lucchini}}, \bibinfo {author} {\bibfnamefont
  {M.}~\bibnamefont {Volkov}}, \bibinfo {author} {\bibfnamefont
  {F.}~\bibnamefont {Schlaepfer}}, \bibinfo {author} {\bibfnamefont
  {L.}~\bibnamefont {Gallmann}}, \bibinfo {author} {\bibfnamefont
  {U.}~\bibnamefont {Keller}}, \ and\ \bibinfo {author} {\bibfnamefont
  {A.}~\bibnamefont {Rubio}},\ }\href {\doibase 10.1103/PhysRevB.98.035202}
  {\bibfield  {journal} {\bibinfo  {journal} {Phys. Rev. B}\ }\textbf {\bibinfo
  {volume} {98}},\ \bibinfo {pages} {035202} (\bibinfo {year}
  {2018})}\BibitemShut {NoStop}%
\end{thebibliography}%


\begin{thebibliography}{7}%
\makeatletter
\providecommand \@ifxundefined [1]{%
 \@ifx{#1\undefined}
}%
\providecommand \@ifnum [1]{%
 \ifnum #1\expandafter \@firstoftwo
 \else \expandafter \@secondoftwo
 \fi
}%
\providecommand \@ifx [1]{%
 \ifx #1\expandafter \@firstoftwo
 \else \expandafter \@secondoftwo
 \fi
}%
\providecommand \natexlab [1]{#1}%
\providecommand \enquote  [1]{``#1''}%
\providecommand \bibnamefont  [1]{#1}%
\providecommand \bibfnamefont [1]{#1}%
\providecommand \citenamefont [1]{#1}%
\providecommand \href@noop [0]{\@secondoftwo}%
\providecommand \href [0]{\begingroup \@sanitize@url \@href}%
\providecommand \@href[1]{\@@startlink{#1}\@@href}%
\providecommand \@@href[1]{\endgroup#1\@@endlink}%
\providecommand \@sanitize@url [0]{\catcode `\\12\catcode `\$12\catcode
  `\&12\catcode `\#12\catcode `\^12\catcode `\_12\catcode `\%12\relax}%
\providecommand \@@startlink[1]{}%
\providecommand \@@endlink[0]{}%
\providecommand \url  [0]{\begingroup\@sanitize@url \@url }%
\providecommand \@url [1]{\endgroup\@href {#1}{\urlprefix }}%
\providecommand \urlprefix  [0]{URL }%
\providecommand \Eprint [0]{\href }%
\providecommand \doibase [0]{http://dx.doi.org/}%
\providecommand \selectlanguage [0]{\@gobble}%
\providecommand \bibinfo  [0]{\@secondoftwo}%
\providecommand \bibfield  [0]{\@secondoftwo}%
\providecommand \translation [1]{[#1]}%
\providecommand \BibitemOpen [0]{}%
\providecommand \bibitemStop [0]{}%
\providecommand \bibitemNoStop [0]{.\EOS\space}%
\providecommand \EOS [0]{\spacefactor3000\relax}%
\providecommand \BibitemShut  [1]{\csname bibitem#1\endcsname}%
\let\auto@bib@innerbib\@empty
\bibitem [{\citenamefont {Sato}\ \emph
  {et~al.}(2019{\natexlab{a}})\citenamefont {Sato}, \citenamefont {McIver},
  \citenamefont {Nuske}, \citenamefont {Tang}, \citenamefont {Jotzu},
  \citenamefont {Schulte}, \citenamefont {H\"ubener}, \citenamefont
  {De~Giovannini}, \citenamefont {Mathey}, \citenamefont {Sentef},
  \citenamefont {Cavalleri},\ and\ \citenamefont {Rubio}}]{PhysRevB.99.214302}%
  \BibitemOpen
  \bibfield  {author} {\bibinfo {author} {\bibfnamefont {S.~A.}\ \bibnamefont
  {Sato}}, \bibinfo {author} {\bibfnamefont {J.~W.}\ \bibnamefont {McIver}},
  \bibinfo {author} {\bibfnamefont {M.}~\bibnamefont {Nuske}}, \bibinfo
  {author} {\bibfnamefont {P.}~\bibnamefont {Tang}}, \bibinfo {author}
  {\bibfnamefont {G.}~\bibnamefont {Jotzu}}, \bibinfo {author} {\bibfnamefont
  {B.}~\bibnamefont {Schulte}}, \bibinfo {author} {\bibfnamefont
  {H.}~\bibnamefont {H\"ubener}}, \bibinfo {author} {\bibfnamefont
  {U.}~\bibnamefont {De~Giovannini}}, \bibinfo {author} {\bibfnamefont
  {L.}~\bibnamefont {Mathey}}, \bibinfo {author} {\bibfnamefont {M.~A.}\
  \bibnamefont {Sentef}}, \bibinfo {author} {\bibfnamefont {A.}~\bibnamefont
  {Cavalleri}}, \ and\ \bibinfo {author} {\bibfnamefont {A.}~\bibnamefont
  {Rubio}},\ }\href {\doibase 10.1103/PhysRevB.99.214302} {\bibfield  {journal}
  {\bibinfo  {journal} {Phys. Rev. B}\ }\textbf {\bibinfo {volume} {99}},\
  \bibinfo {pages} {214302} (\bibinfo {year} {2019}{\natexlab{a}})}\BibitemShut
  {NoStop}%
\bibitem [{\citenamefont {Sato}\ \emph
  {et~al.}(2019{\natexlab{b}})\citenamefont {Sato}, \citenamefont {Tang},
  \citenamefont {Sentef}, \citenamefont {Giovannini}, \citenamefont
  {H\"ubener},\ and\ \citenamefont {Rubio}}]{sato_njp_2019}%
  \BibitemOpen
  \bibfield  {author} {\bibinfo {author} {\bibfnamefont {S.~A.}\ \bibnamefont
  {Sato}}, \bibinfo {author} {\bibfnamefont {P.}~\bibnamefont {Tang}}, \bibinfo
  {author} {\bibfnamefont {M.~A.}\ \bibnamefont {Sentef}}, \bibinfo {author}
  {\bibfnamefont {U.~D.}\ \bibnamefont {Giovannini}}, \bibinfo {author}
  {\bibfnamefont {H.}~\bibnamefont {H\"ubener}}, \ and\ \bibinfo {author}
  {\bibfnamefont {A.}~\bibnamefont {Rubio}},\ }\href {\doibase
  10.1088/1367-2630/ab3acf} {\bibfield  {journal} {\bibinfo  {journal} {New
  Journal of Physics}\ }\textbf {\bibinfo {volume} {21}},\ \bibinfo {pages}
  {093005} (\bibinfo {year} {2019}{\natexlab{b}})}\BibitemShut {NoStop}%
\bibitem [{\citenamefont {Meier}\ \emph {et~al.}(1994)\citenamefont {Meier},
  \citenamefont {von Plessen}, \citenamefont {Thomas},\ and\ \citenamefont
  {Koch}}]{PhysRevLett.73.902}%
  \BibitemOpen
  \bibfield  {author} {\bibinfo {author} {\bibfnamefont {T.}~\bibnamefont
  {Meier}}, \bibinfo {author} {\bibfnamefont {G.}~\bibnamefont {von Plessen}},
  \bibinfo {author} {\bibfnamefont {P.}~\bibnamefont {Thomas}}, \ and\ \bibinfo
  {author} {\bibfnamefont {S.~W.}\ \bibnamefont {Koch}},\ }\href {\doibase
  10.1103/PhysRevLett.73.902} {\bibfield  {journal} {\bibinfo  {journal} {Phys.
  Rev. Lett.}\ }\textbf {\bibinfo {volume} {73}},\ \bibinfo {pages} {902}
  (\bibinfo {year} {1994})}\BibitemShut {NoStop}%
\bibitem [{\citenamefont {Houston}(1940)}]{PhysRev.57.184}%
  \BibitemOpen
  \bibfield  {author} {\bibinfo {author} {\bibfnamefont {W.~V.}\ \bibnamefont
  {Houston}},\ }\href {\doibase 10.1103/PhysRev.57.184} {\bibfield  {journal}
  {\bibinfo  {journal} {Phys. Rev.}\ }\textbf {\bibinfo {volume} {57}},\
  \bibinfo {pages} {184} (\bibinfo {year} {1940})}\BibitemShut {NoStop}%
\bibitem [{\citenamefont {Krieger}\ and\ \citenamefont
  {Iafrate}(1986)}]{PhysRevB.33.5494}%
  \BibitemOpen
  \bibfield  {author} {\bibinfo {author} {\bibfnamefont {J.~B.}\ \bibnamefont
  {Krieger}}\ and\ \bibinfo {author} {\bibfnamefont {G.~J.}\ \bibnamefont
  {Iafrate}},\ }\href {\doibase 10.1103/PhysRevB.33.5494} {\bibfield  {journal}
  {\bibinfo  {journal} {Phys. Rev. B}\ }\textbf {\bibinfo {volume} {33}},\
  \bibinfo {pages} {5494} (\bibinfo {year} {1986})}\BibitemShut {NoStop}%
\bibitem [{\citenamefont {Floss}\ \emph {et~al.}(2018)\citenamefont {Floss},
  \citenamefont {Lemell}, \citenamefont {Wachter}, \citenamefont {Smejkal},
  \citenamefont {Sato}, \citenamefont {Tong}, \citenamefont {Yabana},\ and\
  \citenamefont {Burgd\"orfer}}]{PhysRevA.97.011401}%
  \BibitemOpen
  \bibfield  {author} {\bibinfo {author} {\bibfnamefont {I.}~\bibnamefont
  {Floss}}, \bibinfo {author} {\bibfnamefont {C.}~\bibnamefont {Lemell}},
  \bibinfo {author} {\bibfnamefont {G.}~\bibnamefont {Wachter}}, \bibinfo
  {author} {\bibfnamefont {V.}~\bibnamefont {Smejkal}}, \bibinfo {author}
  {\bibfnamefont {S.~A.}\ \bibnamefont {Sato}}, \bibinfo {author}
  {\bibfnamefont {X.-M.}\ \bibnamefont {Tong}}, \bibinfo {author}
  {\bibfnamefont {K.}~\bibnamefont {Yabana}}, \ and\ \bibinfo {author}
  {\bibfnamefont {J.}~\bibnamefont {Burgd\"orfer}},\ }\href {\doibase
  10.1103/PhysRevA.97.011401} {\bibfield  {journal} {\bibinfo  {journal} {Phys.
  Rev. A}\ }\textbf {\bibinfo {volume} {97}},\ \bibinfo {pages} {011401}
  (\bibinfo {year} {2018})}\BibitemShut {NoStop}%
\bibitem [{\citenamefont {Sato}\ \emph {et~al.}(2018)\citenamefont {Sato},
  \citenamefont {Lucchini}, \citenamefont {Volkov}, \citenamefont {Schlaepfer},
  \citenamefont {Gallmann}, \citenamefont {Keller},\ and\ \citenamefont
  {Rubio}}]{PhysRevB.98.035202}%
  \BibitemOpen
  \bibfield  {author} {\bibinfo {author} {\bibfnamefont {S.~A.}\ \bibnamefont
  {Sato}}, \bibinfo {author} {\bibfnamefont {M.}~\bibnamefont {Lucchini}},
  \bibinfo {author} {\bibfnamefont {M.}~\bibnamefont {Volkov}}, \bibinfo
  {author} {\bibfnamefont {F.}~\bibnamefont {Schlaepfer}}, \bibinfo {author}
  {\bibfnamefont {L.}~\bibnamefont {Gallmann}}, \bibinfo {author}
  {\bibfnamefont {U.}~\bibnamefont {Keller}}, \ and\ \bibinfo {author}
  {\bibfnamefont {A.}~\bibnamefont {Rubio}},\ }\href {\doibase
  10.1103/PhysRevB.98.035202} {\bibfield  {journal} {\bibinfo  {journal} {Phys.
  Rev. B}\ }\textbf {\bibinfo {volume} {98}},\ \bibinfo {pages} {035202}
  (\bibinfo {year} {2018})}\BibitemShut {NoStop}%
\end{thebibliography}%

\end{document}


\author{Shunsuke~A.~Sato}
\email{ssato@ccs.tsukuba.ac.jp}
\affiliation 
{Center for Computational Sciences, University of Tsukuba, Tsukuba 305-8577, Japan}
\affiliation 
{Max Planck Institute for the Structure and Dynamics of Matter, Luruper Chaussee 149, 22761 Hamburg, Germany}

\author{Hideki~Hirori}
\affiliation
{Institute for Chemical Research, Kyoto University, Uji, Kyoto 611-0011, Japan}

\author{Yasuyuki~Sanari}
\affiliation
{Institute for Chemical Research, Kyoto University, Uji, Kyoto 611-0011, Japan}

\author{Yoshihiko~Kanemitsu}
\affiliation
{Institute for Chemical Research, Kyoto University, Uji, Kyoto 611-0011, Japan}

\author{Angel~Rubio}
\affiliation 
{Max Planck Institute for the Structure and Dynamics of Matter, Luruper Chaussee 149, 22761 Hamburg, Germany}
\affiliation 
{Center for Computational Quantum Physics (CCQ), Flatiron Institute, 162 Fifth Avenue, New York, NY
10010, USA}

\title{Supplemental Material: High-order harmonic generation in graphene: nonlinear coupling of intra and interband transitions
}

\maketitle

\section{Equation of motion and relaxation operator}
Here, we describe the details of our theoretical modeling of electron dynamics in graphene under laser fields. The model has been introduced in the previous works \cite{PhysRevB.99.214302,sato_njp_2019}. The electron dynamics is described by the following quantum master equation for the one-body electron density matrix,
\be
\frac{d}{dt} \rho_{\vecb k}(t)=\frac{1}{i\hbar} \left [H_{\vecb k + \vecb A(t)} \right ] + \hat D \left [ \rho_{\vecb k}(t)\right ],
\ee
where $H_{\vecb k + \vecb A(t)}$ is the time-dependent Hamiltonian and $\hat D \left [ \rho_{\vecb k}(t) \right ]$ is a relaxation operator.
In this work, we consider the following $2$-by-$2$ Hamiltonian matrix
\be
H_{\vecb k + \vecb A(t)} = v_F \tau_z \sigma_x \left [ k_x +A_x(t) \right ]
+ v_F \sigma_y \left [k_y + A_y(t) \right ] + \frac{\Delta}{2},
\ee
where $\sigma_j$ are the Pauli matrices.

For the relaxation operator, we employ a simple relaxation time approximation \cite{PhysRevLett.73.902} in the instantaneous eigenbasis expression \cite{PhysRev.57.184,PhysRevB.33.5494}. For this purpose, we first introduce instantaneous eigenstates of the Hamiltonian $H_{\vecb k + \vecb A(t)}$
\be
H_{\vecb k + \vecb A(t)} \left |u_{b,\vecb k + \vecb A(t)} \right \rangle
= \epsilon_{b,\vecb k + \vecb A(t)}
\left |u_{b,\vecb k + \vecb A(t)} \right \rangle,
\ee
where $b$ is a band index, $\left |u_{b,\vecb k + \vecb A(t)} \right >$ are instantaneous eigenstates, and $\epsilon_{b,\vecb k + \vecb A(t)}$ are the corresponding single-particle energies. Note that, since we consider the $2$-by-$2$ Hamiltonian, we have only two bands: one is the valence band ($b=v$), and the other is the conduction band ($b=c$). One can further introduce a unitary matrix $U_{\vecb k + \vecb A(t)}$ as $U_{\vecb k + \vecb A(t)}= \left (\left |u_{v,\vecb k + \vecb A(t)} \right \rangle, \left |u_{c,\vecb k + \vecb A(t)} \right \rangle \right )$. The introduced unitary matrix diagonalizes the Hamiltonian as
\be
U^{\dagger}_{\vecb k + \vecb A(t)} H_{\vecb k+ \vecb A(t)} U_{\vecb k+ \vecb A(t)}
=
\left(
    \begin{array}{cc}
      \epsilon_{v, \vecb k+ \vecb A(t)} & 
      0 \\
      0 &
      \epsilon_{c, \vecb k+ \vecb A(t)}
    \end{array}
  \right).
\ee
Based on the unitary matrix, we further introduce the one-body density matrix in the instantaneous eigenbasis representation as
\be
\tilde \rho_{\vecb k}(t) = \left(
    \begin{array}{cc}
      \rho_{vv,\vecb k}(t) & 
      \rho_{vc,\vecb k}(t) \\
      \rho_{cv,\vecb k}(t) &
      \rho_{cc,\vecb k}(t)
    \end{array}
    \right) = U_{\vecb k + \vecb A(t)}^{\dagger}  \rho_{\vecb k}(t) U_{\vecb k + \vecb A(t)}.
\ee

With the instantaneous eigenbasis representation, we consider the relaxation of the density matrix by introducing two kinds of relaxation: One is the longitudinal relaxation, which is the relaxation of the diagonal elements of the density matrix, while the other is the transverse relaxation, which is the relaxation of the off-diagonal elements. To realize such relaxation in a practical implementation, we introduce the relaxation operator as
\be
\hat D\left [\rho_{\vecb k}(t)\right] =-
U_{\vecb k}(t)
\left(
    \begin{array}{cc}
      \frac{\rho_{vv,\vecb k}(t)-\rho^{F}_{v,\vecb k + \vecb A(t)}}{T_1} & 
      \frac{\rho_{vc,\vecb k}(t)}{T_2} \\
      \frac{\rho_{cv,\vecb k}(t)}{T_2} &
      \frac{\rho_{cc,\vecb k}(t)-\rho^{F}_{c,\vecb k + \vecb A(t)}}{T_1}
    \end{array}
  \right)U^{\dagger}_{\vecb k}(t), \nonumber \\
\label{eq:relaxation-op}
\ee
where $T_1$ and $T_2$ are the time-constants for the longitudinal and transverse relaxations, respectively. In this work, as a target state of the relaxation, we employ the Fermi-Dirac distribution
\be
\rho^F_{b,\vecb k}=\frac{1}{e^{(\epsilon_{b,\vecb k}-\mu)/k_BT_e}+1},
\ee
where $\mu$ is the chemical potential, $T_e$ is the electron temperature.
In this work, we fix the electron temperature $T_e$ to $80$~K.

\section{Focal spot average \label{subsec:focal_average}}

In realistic experimental configurations, the high-order harmonic generation occurs not only at the center of a beam-spot but also on the whole focal area. Thus, the generated high-order harmonics from a wide region of the focal area can be detected. To take into account the macroscopic focal-spot average effect of HHG, we employ the intensity average procedure according to Ref.~\cite{PhysRevA.97.011401}.

Here, we assume the field strength of laser electric fields has the following Gaussian profile on the sample surface
\be
E(x,y)=E_0 \exp \left [ -\frac{1}{\sigma^2} \left (x^2+y^2 \right )\right ],
\ee
where $(x,y)$ are the coordinates on the surface, $E_0$ is the peak field strength, and $\sigma$ is the beam waist. We further assume that the beam waist is sufficiently large, and the induced current depends only on the local field strength as $\vecb J[E(x,y)](t)$. Based on these assumptions, the average current within the beam spot can be evaluated as
\be
\vecb J^{ave}(t) = \frac{1}{\pi \sigma^2} \int dx dy \vecb J[E(x,y)]
= \int^1_0 d \alpha \frac{1}{\alpha} \vecb J[\alpha E_0](t).
\label{eq:focal-average}
\ee
Hence the total current on the sample can be evaluated as the intensity average of the induced current. In this work, we repeatedly perform the simulation by changing the laser field strength $E_0 =\sqrt{E^2_{0,x}+E^2_{0,y}}$ with fixed ratio of $E_{0,x}$ and $E_{0,y}$. Then, we compute the averaged current on the sample $\vecb J^{ave}(t)$ with Eq.~(\ref{eq:focal-average}).

\section{Decomposition of intraband and interband transitions}

In this work, we define intraband and interband transitions based on the instantaneous eigenbasis representation \cite{PhysRev.57.184,PhysRevB.33.5494}. To define the intraband and interband transitions, we first consider the equation of motion of the reduced density matrix in the instantaneous eigenbasis representation as
\be
\frac{d}{dt}\tilde \rho_{\vecb k}(t)
&=&\frac{d}{dt}
U_{\vecb k + \vecb A(t)}^{\dagger}  \rho_{\vecb k}(t) U_{\vecb k + \vecb A(t)}
=
\frac{dU_{\vecb k + \vecb A(t)}^{\dagger}}{dt}
  \rho_{\vecb k}(t) U_{\vecb k + \vecb A(t)}
+ U_{\vecb k + \vecb A(t)}^{\dagger}  \frac{d\rho_{\vecb k}(t)}{dt} U_{\vecb k + \vecb A(t)}
+ U_{\vecb k + \vecb A(t)}^{\dagger}  \rho_{\vecb k}(t) \frac{dU_{\vecb k + \vecb A(t)}}{dt}
\nonumber \\
&=&\frac{1}{i\hbar} \left [\tilde H_{\vecb k + \vecb A(t)},
\tilde  \rho_{\vecb k}(t)
\right ] 
- \left(
    \begin{array}{cc}
      \frac{\rho_{vv,\vecb k}(t)-\rho^{F}_{v,\vecb k+\vecb A(t)}}{T_1} & 
      \frac{\rho_{vc,\vecb k}(t)}{T_2} \\
      \frac{\rho_{cv,\vecb k}(t)}{T_2} &
      \frac{\rho_{cc,\vecb k}(t)-\rho^{F}_{c,\vecb k+\vecb A(t)}}{T_1}
    \end{array}
  \right),
\ee
where the effective Hamiltonian $\tilde H_{\vecb k + \vecb A(t)}$ in the instantaneous eigenbasis representation is given by
\be
\tilde H_{\vecb k + \vecb A(t)} =
\left(
    \begin{array}{cc}
      \epsilon_{v,\vecb k + \vecb A(t)} &
      0 \\
      0 &
      \epsilon_{c,\vecb k + \vecb A(t)} 
    \end{array}
  \right)
  + i\hbar 
\left(
    \begin{array}{cc}
     \big \langle \frac{\partial u_{v,\vecb k + \vecb A(t)}}{\partial \vecb k} \big | u_{v,\vecb k + \vecb A(t)}  \big \rangle  \cdot \dot{\vecb{A}}(t), &
     \big \langle \frac{\partial u_{v,\vecb k + \vecb A(t)}}{\partial \vecb k} \big | u_{c,\vecb k + \vecb A(t)}  \big \rangle  \cdot \dot{\vecb{A}}(t) \\
      \big \langle \frac{\partial u_{c,\vecb k + \vecb A(t)}}{\partial \vecb k} \big | u_{v,\vecb k + \vecb A(t)}  \big \rangle  \cdot \dot{\vecb{A}}(t), &
      \big \langle \frac{\partial u_{c,\vecb k + \vecb A(t)}}{\partial \vecb k} \big | u_{c,\vecb k + \vecb A(t)}  \big \rangle  \cdot \dot{\vecb{A}}(t)
    \end{array}
  \right).
\label{si-eq:ham-eig}
\ee
To naturally distinguish the intraband and interband transitions, we rewrite the Hamiltonian with the new gauge fields, $\vecb A_{tra}$ and $\vecb A_{ter}(t)$ as
\be
\tilde H^\mathrm{tra-ter}_{\vecb k + \vecb A(t)} =
\left(
    \begin{array}{cc}
      \epsilon_{v,\vecb k + \vecb A_{tra}(t)} &
      0 \\
      0 &
      \epsilon_{c,\vecb k + \vecb A_{tra}(t)} 
    \end{array}
  \right)
  + i\hbar 
\left(
    \begin{array}{cc}
     \big \langle \frac{\partial u_{v,\vecb k + \vecb A_{tra}(t)}}{\partial \vecb k} \big | u_{v,\vecb k + \vecb A_{tra}(t)}  \big \rangle  \cdot \dot{\vecb{A}}_{tra}(t), &
     \big \langle \frac{\partial u_{v,\vecb k + \vecb A_{tra}(t)}}{\partial \vecb k} \big | u_{c,\vecb k + \vecb A_{tra}(t)}  \big \rangle  \cdot \dot{\vecb{A}}_{ter}(t) \\
      \big \langle \frac{\partial u_{c,\vecb k + \vecb A_{tra}(t)}}{\partial \vecb k} \big | u_{v,\vecb k + \vecb A_{tra}(t)}  \big \rangle  \cdot \dot{\vecb{A}}_{ter}(t), &
      \big \langle \frac{\partial u_{c,\vecb k + \vecb A_{tra}(t)}}{\partial \vecb k} \big | u_{c,\vecb k + \vecb A_{tra}(t)}  \big \rangle  \cdot \dot{\vecb{A}}_{tra}(t)
    \end{array}
  \right). \nonumber \\
\label{si-eq:ham-eig-intra-inter}
\ee
Note that, if $\vecb A_{tra}(t) = \vecb A_{ter}(t)=\vecb A(t)$, the new Hamiltonian $\tilde H^\mathrm{tra-ter}_{\vecb k + \vecb A(t)}$ in Eq.~(\ref{si-eq:ham-eig-intra-inter}) is identical to the original Hamiltonian $\tilde H_{\vecb k + \vecb A(t)}$ in Eq.~(\ref{si-eq:ham-eig}). If $\vecb A_{tra}(t)= \vecb A(t)$ and $\vecb A_{ter}(t)=0$, the system is adiabatically evolved within the same band. Hence the transitions induced by $\vecb A_{tra}(t)$ is nothing but the intraband transitions. On the other hand, if $\vecb A_{tra}(t)= 0$ and $\vecb A_{ter}(t)=\vecb A(t)$, the system shows only a transition among different bands without the motion in the $k$-space. Hence the transitions induced by $\vecb{A}_{ter}(t)$ is nothing but the interband transitions. Note that these definitions of the intraband and interband transitions with the reduced density matrix are equivalent to those defined with the Houston expansion in the wavefunction theory \cite{PhysRevB.98.035202}.

Based on the electron dynamics simulations with the effective Hamiltonian $\tilde H^\mathrm{tra-ter}_{\vecb k + \vecb A(t)}$, one can elucidate impacts of intraband and interband transitions. For example, by setting $\vecb A_{tra}(t)$ to $\vecb A(t)$ and $\vecb A_{ter}(t)$ to zero, one can study the electron dynamics induced solely by the intraband transitions. Likewise, by setting $\vecb A_{tra}(t)$ to zero and $\vecb A_{ter}(t)$ to $\vecb A(t)$, one can study the electron dynamics induced solely by the interband transitions. Applying this approach to the HHG in graphene, we evaluated the high-order harmonic intensity induced solely by intraband or interband transitions. The results are shown in Fig.~2~(e) in the main text.

To investigate detailed roles of nonlinear coupling among intraband and interband transitions, we introduce a decomposition of the current $\vecb J(t,E_{0,x},E_{0,y})$ into each nonlinear coupling component. In the above analysis, we introduced the two gauge fields, $\vecb A_{tra}(t)$ and $\vecb A_{ter}(t)$, to distinguish the intraband and interband transitions. Furthermore, each gauge field vector consists of two directional components. Hence, the induced current is a function of the field strength of each component of each gauge field as $\vecb J^\mathrm{tra-ter}(t,E_{0,x,tra},E_{0,x,ter},E_{0,y,tra},E_{0,y,ter})$. In the main text, we introduced the following four labels for each kind of transitions. $\tau_a$: the intraband transitions induced by the $x$-component of fields. $\tau_b$: the interband transitions induced by the $x$-component of fields. $\tau_c$: the intraband transitions induced by the $y$-component of fields. $\tau_d$: the interband transitions induced by the $y$-component of fields. With this notation, we first define four kinds of current as follows:
\be
\vecb J_{\tau_a}(t) &=& \vecb J^\mathrm{tra-ter}(t,E_{0,x,tra},0,0,0), \label{si-eq:j_x_non}\\
\vecb J_{\tau_b}(t) &=& \vecb J^\mathrm{tra-ter}(t,0,E_{0,x,ter},0,0), \label{si-eq:j_non_x}\\
\vecb J_{\tau_c}(t) &=& \vecb J^\mathrm{tra-ter}(t,0,0,E_{0,y,tra},0), \label{si-eq:j_y_non}\\
\vecb J_{\tau_d}(t) &=& \vecb J^\mathrm{tra-ter}(t,0,0,0,E_{0,y,ter}). \label{si-eq:j_non_y}
\ee
Each of them corresponds to the current induced solely by a single-directional component of each transition: $\vecb J_{\tau_a}(t)$ is induced solely by the $x$-component of intraband transitions, and $\vecb J_{\tau_b}(t)$ is induced solely by the $x$-component of inter transitions. Likewise, $\vecb J_{\tau_c}(t)$ is induced solely by the $y$-component of intraband transitions, and $\vecb J_{\tau_d}(t)$ is induced solely by the $y$-component of interband transitions.

Then, we define the current induced by nonlinear coupling of two of four components as
\be
\vecb J_{\tau_a, \tau_b}(t) &=& \vecb J^\mathrm{tra-ter}(t,E_{0,x,tra},E_{0,x,ter},0,0,)-\vecb J_{\tau_a}(t)-\vecb J_{\tau_b}(t) \label{si-eq:j_x_x} \\
\vecb J_{\tau_a, \tau_c}(t) &=& \vecb J^\mathrm{tra-ter}(t,E_{0,x,tra},0,E_{0,y,tra},0)-\vecb J_{\tau_a}(t)-\vecb J_{\tau_c}(t) \label{si-eq:j_xy_non} \\
\vecb J_{\tau_a, \tau_d}(t) &=& \vecb J^\mathrm{tra-ter}(t,E_{0,x,tra},0,0,E_{0,y,ter})-\vecb J_{\tau_a}(t)-\vecb J_{\tau_d}(t) \label{si-eq:j_x_y} \\
\vecb J_{\tau_b, \tau_c}(t) &=& \vecb J^\mathrm{tra-ter}(t,0,E_{0,x,ter},E_{0,y,tra},0)-\vecb J_{\tau_b}(t)-\vecb J_{\tau_c}(t) \label{si-eq:j_y_x} \\
\vecb J_{\tau_b, \tau_d}(t) &=& \vecb J^\mathrm{tra-ter}(t,0,E_{0,x,ter},0,E_{0,y,ter})-\vecb J_{\tau_b}(t)-\vecb J_{\tau_d}(t) \label{si-eq:j_non_xy} \\
\vecb J_{\tau_c, \tau_d}(t) &=& \vecb J^\mathrm{tra-ter}(t,0,0,E_{0,y,tra},E_{0,y,ter})-\vecb J_{\tau_c}(t)-\vecb J_{\tau_d}(t).  \label{si-eq:j_y_y}
\ee
Here, $\vecb J_{\tau_a,\tau_b}(t)$ is induced by the nonlinear coupling of the $x$-components of the intraband and interband transitions, $\vecb J_{\tau_a,\tau_c}(t)$ is induced by the nonlinear coupling of the $x$- and $y$-components of the intraband transitions, $\vecb J_{\tau_a,\tau_d}(t)$ is induced by the nonlinear coupling of the $x$-component of the intraband and the $y$-component of the interband transitions, $\vecb J_{\tau_b,\tau_c}(t)$ is induced by the nonlinear coupling of the $x$-component of the interband and the $y$-component of the intraband transitions, $\vecb J_{\tau_b,\tau_d}(t)$ is induced by the nonlinear coupling of the $x$- and $y$-components of the interband transitions, and $\vecb J_{\tau_c,\tau_d}(t)$ is induced by the nonlinear coupling of the $y$-components of the intraband and intraband transitions.

In the same way, we define the current induced by nonlinear coupling among three of four field components as
\be
\vecb J_{\tau_a, \tau_b, \tau_c}(t) &=& \vecb J^\mathrm{tra-ter}(t,E_{0,x,tra},E_{0,x,ter},E_{0,y,tra},0)-\vecb J_{\tau_a, \tau_b}(t)-\vecb J_{\tau_a, \tau_c}(t)-\vecb J_{\tau_b, \tau_c}(t) \label{si-eq:j_xy_x} \nonumber \\
&& -\vecb J_{\tau_a}(t)-\vecb J_{\tau_b}(t)-\vecb J_{\tau_c}(t) \\
\vecb J_{\tau_a, \tau_b, \tau_d}(t) &=& \vecb J^\mathrm{tra-ter}(t,E_{0,x,tra},E_{0,x,ter},0,E_{0,y,ter})-\vecb J_{\tau_a, \tau_b}(t)-\vecb J_{\tau_a, \tau_d}(t)-\vecb J_{\tau_b, \tau_d}(t) \nonumber \\
&& -\vecb J_{\tau_a}(t)-\vecb J_{\tau_b}(t)-\vecb J_{\tau_d}(t) \\
\vecb J_{\tau_a, \tau_c, \tau_d}(t) &=& \vecb J^\mathrm{tra-ter}(t,E_{0,x,tra},0,E_{0,y,tra},E_{0,y,ter})-\vecb J_{\tau_a, \tau_c}(t)-\vecb J_{\tau_a, \tau_d}(t)-\vecb J_{\tau_c, \tau_d}(t) \nonumber \\
&& -\vecb J_{\tau_a}(t)-\vecb J_{\tau_c}(t)-\vecb J_{\tau_d}(t) \\
\vecb J_{\tau_b, \tau_c, \tau_d}(t) &=& \vecb J^\mathrm{tra-ter}(t,0,E_{0,x,ter},E_{0,y,tra},E_{0,y,ter})-\vecb J_{\tau_b, \tau_c}(t)-\vecb J_{\tau_b, \tau_d}(t)-\vecb J_{\tau_c, \tau_d}(t) \nonumber \\
&& -\vecb J_{\tau_b}(t)-\vecb J_{\tau_c}(t)-\vecb J_{\tau_d}(t). \label{si-eq:j_y_xy}
\ee

Finally, we define the current induced by nonlinear coupling of all of the field components as
\be
\vecb J_{\tau_a, \tau_b, \tau_c, \tau_d}(t) &=& \vecb J^\mathrm{tra-ter}(t,E_{0,x,tra},E_{0,x,ter},E_{0,y,tra},E_{0,y,ter}) -\vecb J_{\tau_a, \tau_b, \tau_c}(t) -\vecb J_{\tau_a, \tau_b, \tau_d}(t) -\vecb J_{\tau_a, \tau_c, \tau_d}(t) -\vecb J_{\tau_b, \tau_c, \tau_d}(t) \nonumber \\
&-& \vecb J_{\tau_a, \tau_b}(t) -\vecb J_{\tau_a, \tau_c}(t) -\vecb J_{\tau_a, \tau_d}(t) -\vecb J_{\tau_b, \tau_c}(t) -\vecb J_{\tau_b, \tau_d}(t) -\vecb J_{\tau_c, \tau_d}(t) \nonumber \\
&-& \vecb J_{\tau_a}(t) -\vecb J_{\tau_b}(t)-\vecb J_{\tau_c}(t)-\vecb J_{\tau_d}(t).
\label{si-eq:j_xy_xy}
\ee

By construction of the decomposed current in Eqs.~(\ref{si-eq:j_x_non}-\ref{si-eq:j_xy_xy}), the total current $\vecb J(t,E_{0,x},E_{0,y})$ is fully reconstructed as
\be
\vecb J(t,E_{0,x},E_{0,y}) &=& 
\vecb J_{\tau_a}(t) + \vecb J_{\tau_b}(t)
\vecb J_{\tau_c}(t) + \vecb J_{\tau_d}(t) \nonumber \\
&+& \vecb J_{\tau_a, \tau_b}(t) + \vecb J_{\tau_a, \tau_c}(t) +\vecb J_{\tau_a, \tau_d}(t)
+\vecb J_{\tau_b, \tau_c}(t)+\vecb J_{\tau_b, \tau_d}(t)+\vecb J_{\tau_c, \tau_d}(t) \nonumber \\
&+&\vecb J_{\tau_a, \tau_b, \tau_c}(t)+\vecb J_{\tau_a, \tau_b, \tau_d}(t)+\vecb J_{\tau_a, \tau_c, \tau_d}(t)+\vecb J_{\tau_b, \tau_c, \tau_d}(t)
\nonumber \\
&+&\vecb J_{\tau_a, \tau_b, \tau_c, \tau_d}(t).
\ee

\begin{figure}[htbp]
  \includegraphics[width=0.5\columnwidth]{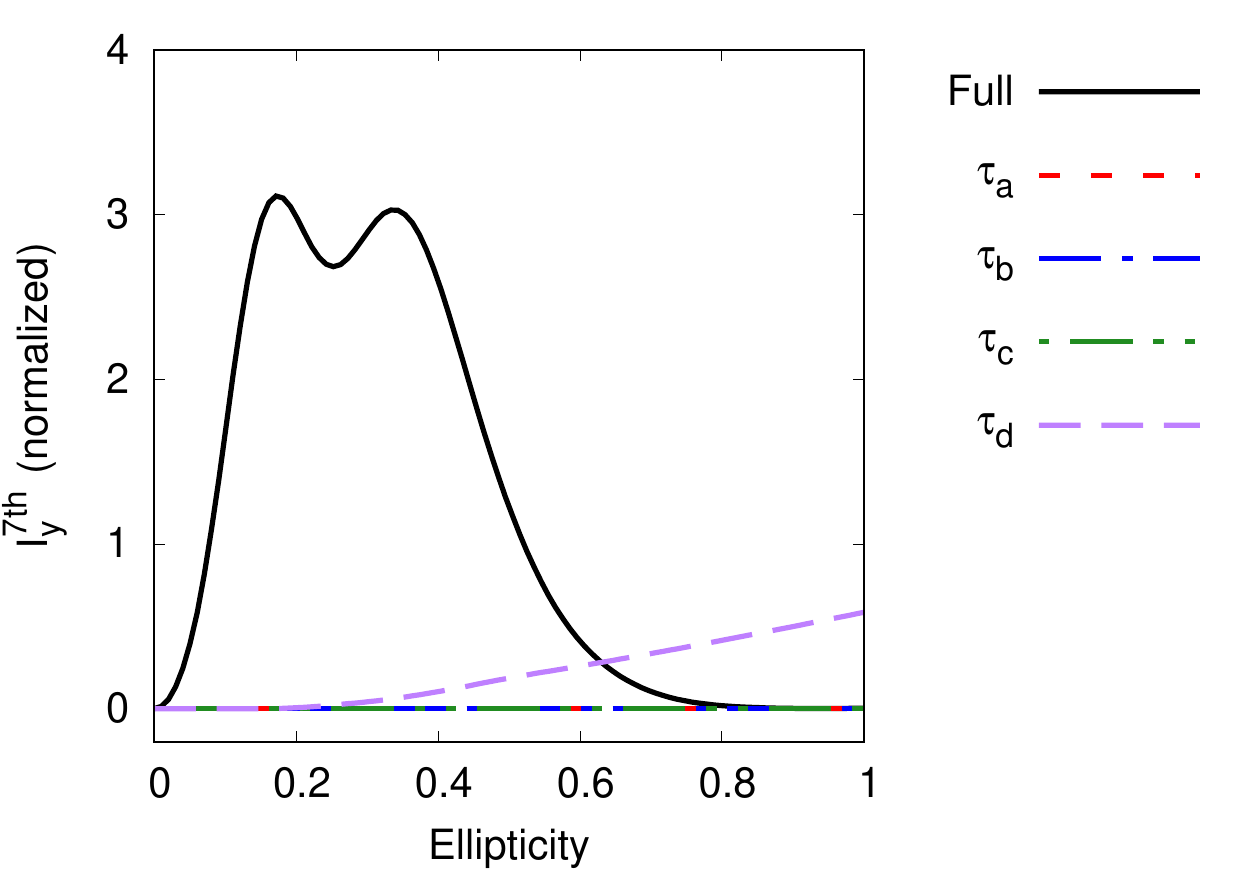}
\caption{\label{fig:SI_intra_inter_hhg_7th_single}
The 7th-order HHG from graphene under elliptically polarized light. Here, the contributions from each single transition $\vecb J_{\tau}$ in Eqs.~(\ref{si-eq:j_x_non}-\ref{si-eq:j_non_y}) are shown.
}
\end{figure}

\begin{figure}[htbp]
  \includegraphics[width=0.5\columnwidth]{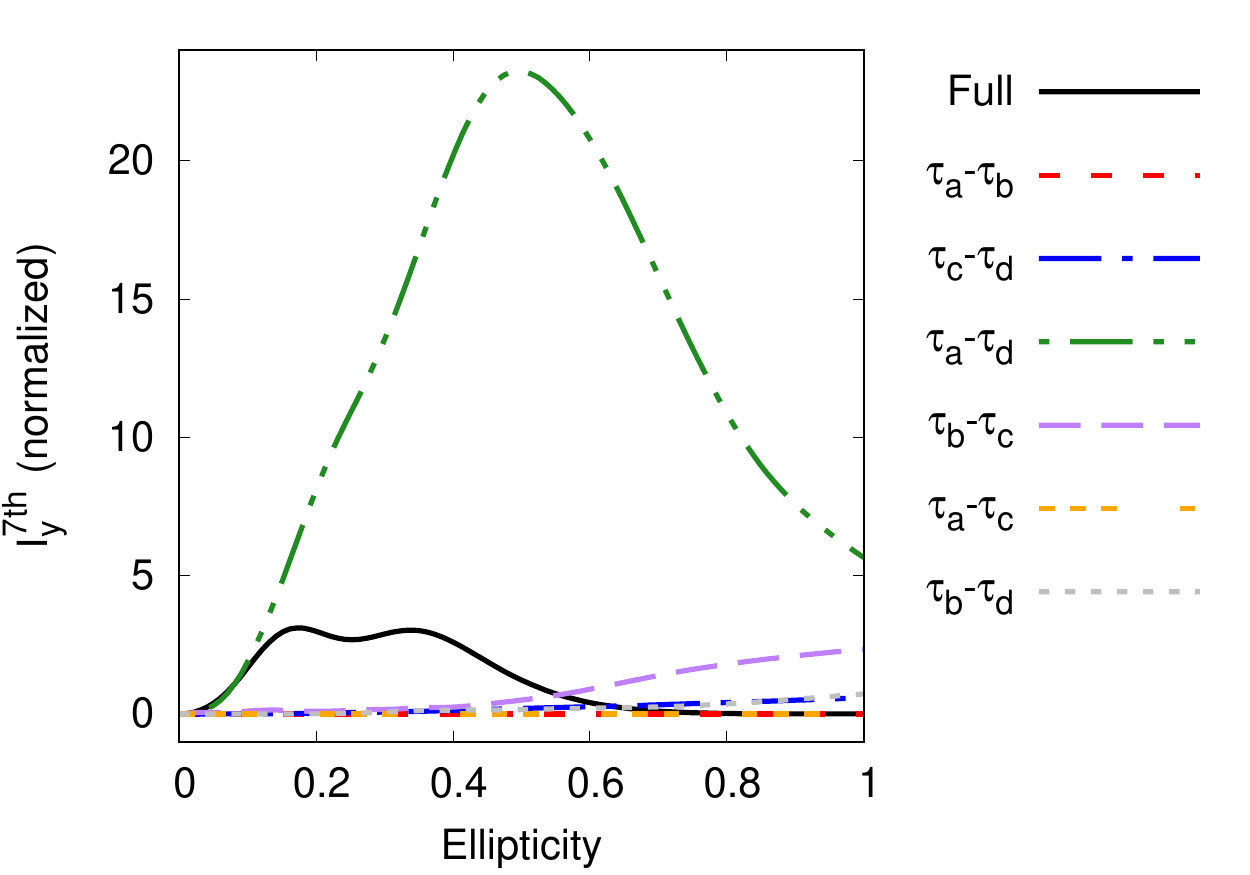}
\caption{\label{fig:SI_intra_inter_hhg_7th_double}
The 7th-order HHG from graphene under elliptically polarized light. Here, the contributions from nonlinear coupling of two of four transitions $\vecb J_{\tau \sigma}$ in Eqs.~(\ref{si-eq:j_x_non}-\ref{si-eq:j_non_y}) are shown.
}
\end{figure}

\begin{figure}[htbp]
  \includegraphics[width=0.5\columnwidth]{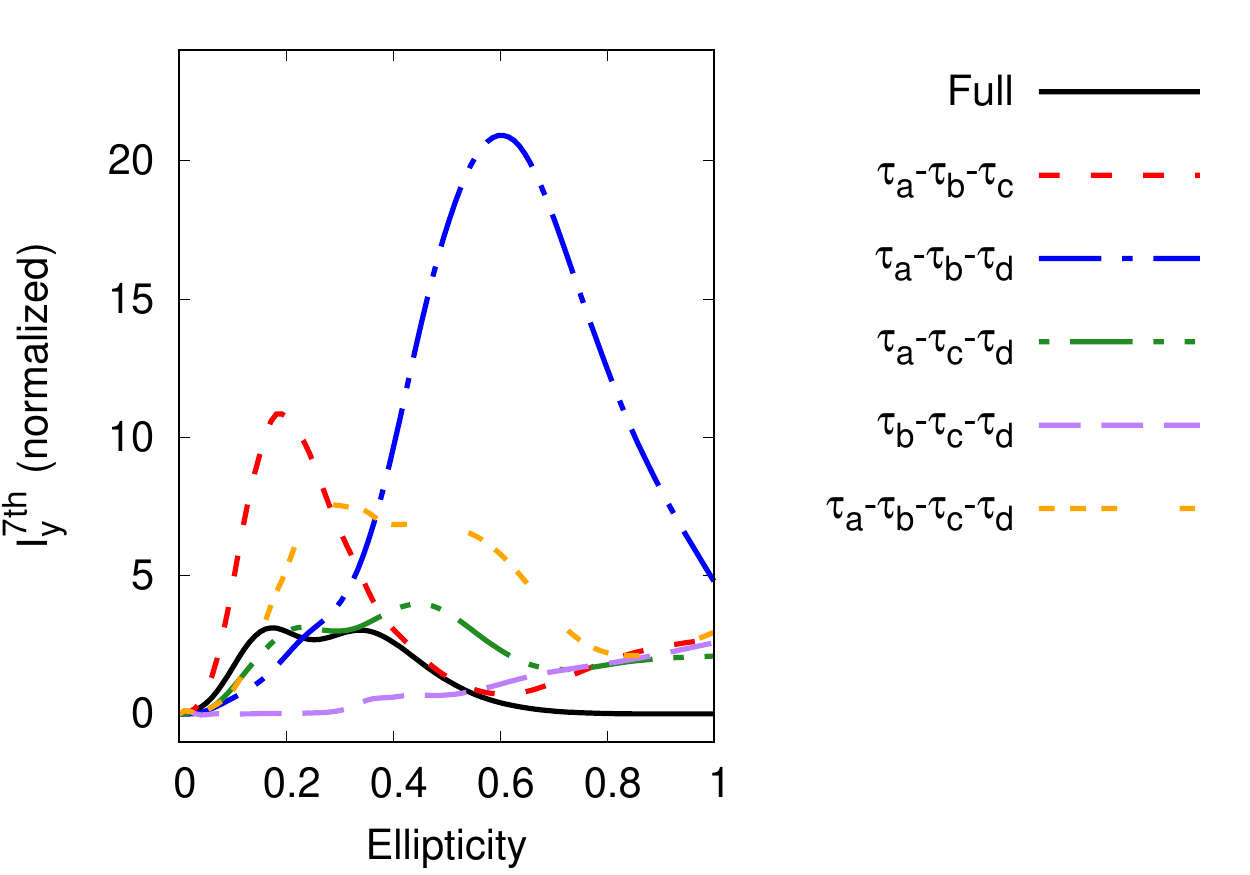}
\caption{\label{fig:SI_intra_inter_hhg_7th_triple_all}
The 7th-order HHG from graphene under elliptically polarized light. Here, the contributions from nonlinear coupling of three of four transitions $\vecb J_{\tau \sigma \delta}$ in Eqs.~(\ref{si-eq:j_xy_x}-\ref{si-eq:j_y_xy}) are shown. The contribution from the nonlinear coupling of all of the transitions $\vecb J_{\tau_a, \tau_b, \tau_c, \tau_d}(t)$ in Eq.~(\ref{si-eq:j_xy_xy}) is also shown.
}
\end{figure}

We evaluated the harmonic intensity with each decomposed current. Figures~S\ref{fig:SI_intra_inter_hhg_7th_single}-S\ref{fig:SI_intra_inter_hhg_7th_triple_all} show the $7$th-order harmonic intensity $I^{7th}_y$ as a function of ellipticity for various decomposed current. Here, we employed the same conditions as those of Fig.~2~(f) of the main text. In Fig.~S\ref{fig:SI_intra_inter_hhg_7th_single}, the results of the current induced sorely by a single transition, $\vecb J_{\tau}$, are shown. In Fig.~S\ref{fig:SI_intra_inter_hhg_7th_double}, the results of the current induced by the nonlinear coupling between two of four kinds of transitions, $\vecb J_{\tau \sigma}$, are shown. In Fig.~S\ref{fig:SI_intra_inter_hhg_7th_triple_all}, the results of the current induced by the nonlinear coupling among three of four transitions, $\vecb J_{\tau \sigma \delta}$, and among all of four transitions, $\vecb J_{\tau_a, \tau_b, \tau_c, \tau_d}(t)$, are shown. In these figures, the result of the full transitions is also shown as the black-solid line.

\bibliography{ref}